\documentclass[%
 sor,
 jor,
 amsmath,amssymb,
 reprint,%
]{revtex4-2}

\usepackage{upgreek}
\usepackage{amsmath}

\newcounter{inlineenum}
\renewcommand{\theinlineenum}{\roman{inlineenum}}
\newenvironment{inlineenum}
{\unskip\ignorespaces\setcounter{inlineenum}{0}%
	\renewcommand{\item}{\refstepcounter{inlineenum}{\textit{\theinlineenum})~}}}
{\ignorespacesafterend}

\newcommand{\pa}{\textit{P. aeruginosa }}
\newcommand{\pafull}{\textit{Pseudomonas aeruginosa }}

\newcommand{\cn}{\textit{C. neoformans }}
\newcommand{\cnfull}{\textit{Cryptococcus neoformans }}

\newcommand{\ca}{\textit{C. albicans }}
\newcommand{\cafull}{\textit{Candida albicans }}

\usepackage{graphicx}
\usepackage{dcolumn}
\usepackage{bm}

\begin{document}

\preprint{AIP/123-QED}

\title[Bacterial spreading in a lawn of yeast]{A dynamic fluid landscape mediates the spread of bacteria}

\author{Divakar Badal}
    \email{divakarbadal@iisc.ac.in}
    \affiliation{Department of Bioengineering, Indian Institute of Science, Bengaluru, India}
\author{Aloke Kumar}%
    \affiliation{Department of Mechanical Engineering, Indian Institute of Science, Bengaluru, India}

\author{Varsha Singh}
    \affiliation{Department of Bioengineering, Indian Institute of Science, Bengaluru, India} 
    \affiliation{Department of Developmental Biology and Genetics, Indian Institute of Science, Bengaluru, India}

\author{Danny Raj M}
    \email{dannym@iisc.ac.in}
    \homepage{https://www.dannyraj.com/}
    \affiliation{Department of Chemical Engineering, Indian Institute of Science, Bengaluru, India} 
\date{\today}

\begin{abstract}
Microbial interactions regulate their spread and survival in competitive environments. It is not clear if the physical parameters of the environment regulate the outcome of these interactions. 
In this work, we show that the opportunistic pathogen \pafull occupies a larger area on the substratum in the presence of yeast such as \cnfull\!\!, than without it. At the microscopic level, bacterial cells show an enhanced activity in the vicinity of yeast cells. 
We observe this behaviour even when the live yeast cells are replaced with heat-killed cells or with spherical glass beads of similar morphology, which suggests that the observed behaviour is not specific to the biology of microbes.
Upon careful investigation, we find that a fluid pool is formed around yeast cells which facilitates the swimming of the flagellated \pa\!\!, causing their enhanced motility.
Using mathematical modeling we demonstrate how this local enhancement of bacterial motility leads to the enhanced spread observed at the level of the plate.
We find that the dynamics of the fluid landscape around the bacteria, mediated by the growing yeast lawn, affects the spreading.
For instance, when the yeast lawn grows faster, a bacterial colony prefers a lower initial loading of yeast cells for optimum enhancement in the spread.
We confirm our predictions using \cafull and \cn\!\!, at different initial compositions.
In summary, our work shows the importance of considering the dynamically changing physical environment while studying bacterial motility in complex environments.
\end{abstract}

\keywords{flagellated bacteria, enhanced motility, fluid layer, dynamic neighbourhood, mathematical modelling} 

\maketitle

\begin{quotation}
In what manner does \textit{Pseudomonas aeruginosa} engage in symbiotic relationships with other microorganisms and facilitate its territorial expansion? Our research has shown that \textit{Pseudomonas aeruginosa} uses the naturally formed fluid film surrounding adjacent microorganisms as a means to enhance its dissemination by swimming. This fluid film is dynamic, changing in response to the evolving landscape as the adjacent cells grow and spread.
This study provides insights into the mechanisms by which \textit{Pseudomonas aeruginosa} utilizes physical features to facilitate its spread and colonization within a dynamic microbial ecosystem.\vspace{-10pt}
\end{quotation}

The capacity of microorganisms to thrive in diverse environments lies in their ability to obtain nutrition \cite{Hibbing2010,Zengler2018}. Organisms that locomote have a distinct advantage in comparison to those that do not \cite{Kelly1988,Lauffenburger1991,Bubendorfer2014}. For instance, motile bacteria have a distinct advantage over non-motile microbes in allowing the former to claim a larger fraction of the underlying substratum for growth and proliferation \cite{Conrad2011, Harshey2003}. Most motile bacteria in nature use flagella to propel through a fluid medium rapidly \cite{Reimer2021, Thormann2022}. This necessitates a fluid medium for movement\cite{Wadhwa2022,Purcell1977}. 
However, flagellated bacteria are often found on solid surfaces \cite{Reimer2021,Araujo2019,Belas2013,Potratz2011,HERSHEY2021}; suggesting they either forgo swimming or find sources of fluid to facilitate it.
At least one report suggests that \textit{Escherichia coli} bacteria residing and growing on solid surfaces could form fluid film beneath them when placed on soft agar (0.5\%) surfaces \cite{Wu2012}. This is a condition that facilitates swarming in \textit{E. coli} and \textit{P. aeruginosa} \cite{Kollaran2019,Kearns2010,Zhang2010}. However, bacteria are often found on less fluid-rich environments similar to what is grown in laboratories at higher agar concentrations of about 1-2\%. In such cases, it is not clear how bacteria still exhibit active motility. This implies that the physicochemical elements present in the environment may have a crucial role in shaping locomotion, as well as the subsequent dispersion, nutrient acquisition, and overall success of a population. 

In natural environments, it is unusual to encounter bacteria existing as a solitary species. The microorganisms in their habitat include many types of microbes, such as bacteria and yeast, some of which may also be motile~\cite{Konopka2009,Flemming2016}. Numerous microorganisms exhibit distinct behavioral patterns when exposed to adjacent microorganisms~\cite{Turner1996,Kerr2002,Limoli2019,Trejo-Hernandez2014,Deveau2018,Moran2000,Pradhan2022}. In some cases the changes in behavior are caused by specific chemical signals secreted by these microbes~\cite{Limoli2019,Pradhan2022,Wadhams2004}.
In contrast, Araujo et al~\cite{Araujo2019} showed that the mere presence of graphite particles on agar surface could lead to the formation of fluid film around the particles where motile microbes can exhibit active motility.
Hence, it is plausible to consider that the existence of microorganisms, even those that are immotile, could potentially modify the physical characteristics of their surroundings resulting in modified behavior and motility of nearby microorganisms. 

In this study, we set out to understand how co-existence offers advantages for motile microbes via non-specific cues and/or alteration of the environment. 
We chose \pafull (PA14), a ubiquitously present flagellated bacterium, as the model system for the motile microbe~\cite{Liberati2006}. 
We tested its growth and spread in the presence of a lawn of a model non-motile microbe, the yeast \cnfull (H99$\alpha$) \cite{Kozubowski2012}, on a 1\% agar medium.
We found that \pa spreads better in the presence of \cn than when it is on its own. Surprisingly, we found that the reason for the enhanced spread is not biological in origin, which contrasts with what is commonly believed in the field. A microscopic view of the spreading phenomenon showed that \pa begins to move faster when they were in proximity to \cn microcolonies. Careful experiments showed that this behavior was due to the initiation of swimming by the flagellated bacteria exhibited in a small fluid pool that accumulates around the growing yeast colony. Using a spatially explicit population model for the growing microbes, we showed that this enhancement in motility near the yeast cells gives rise to increased spread of the flagellated bacteria. 
We found that the spreading phenomenon depends on the growth rate ratios and the initial seeding numbers of the microbes. Our model predicted that faster-growing yeast cells increased the spread of motile bacteria in lower concentrations, but inhibited the spread at higher concentrations.
We conducted similar experiments as before with different dilutions of \cn and a faster growing \cafull (SC5314). The spread of \pa observed matched well with the predictions of the model which only considers the growth and spread of microbes due to physical factors.
Our study provides evidence that interaction between organisms need not always be chemical in nature. Non-specific changes in the physico-chemical landscape of the environment mediated by the presence of microorganisms can impact the relative growth and spread. Taken together, this study uncovers the importance of non-motile neighbors in the spread of motile bacteria. 

\section*{Result and Discussion}

\begin{figure}
    \centering
    \includegraphics[width=1\linewidth]{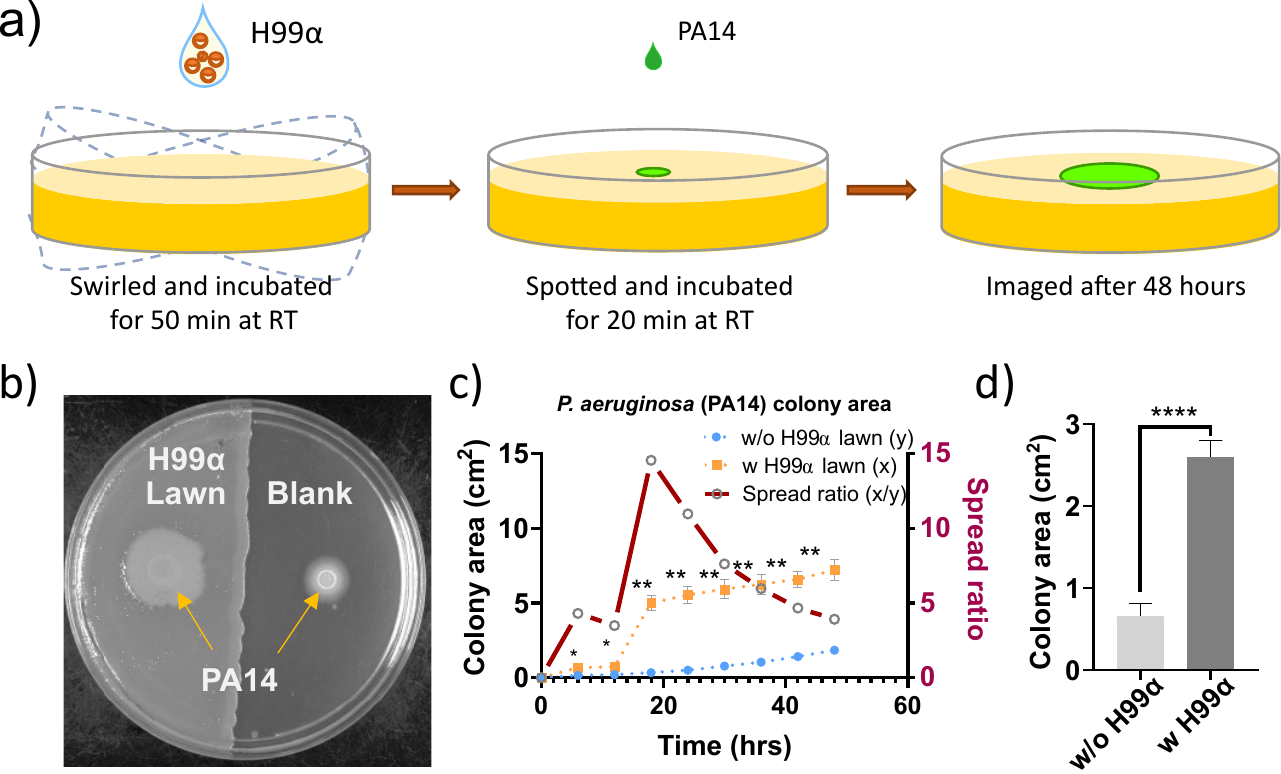}
    \caption{\textbf{\pa shows enhanced spread on a \cn lawn.} (a) Steps involved in preparing the \pa – \cn interaction assay. (b) Snapshots of the spread of \pa\!\!, after 48 hours, on a 90 mm BHI petri dish with only the left half-part covered by the lawn of \cn\!\!. (c) Area of \pa colonies with (orange dotted line) and without (blue dotted line) the lawn of \cn plotted with time, along with the ratio of the areas (enhancement) represented by the red line. Images were recorded every 6 hours for a total of 48 hours. The error bar are based on the standard error. (d) Area of \pa with and without the \cn lawn after 48 hours of incubation. The Student’s t-test with Welch’s correction was used as a statistical test. The p values are depicted by: * p$<$0.05; ** p$<$0.01; *** p$<$0.001; **** p$<$0.0001.}
    \label{fig:enhancedspreadofPAonCNlawn}
\end{figure}

\subsection*{\pa colony spreads faster with \cn\!\!}
To understand how \pa (PA14) spreads in the presence of \cn (H99$\alpha$), we prepared a dynamically growing lawn of \cn by swirling a culture containing these cells (1e9 cells/ml) on a 1\% 90 mm BHI agar plate. Then, we spotted a small volume of \pa culture ($OD_{600}$ 1.5) at the center of the lawn, as illustrated in Figure \ref{fig:enhancedspreadofPAonCNlawn}a. For the purpose of a control run, the same volume of \pa was spotted on a plain BHI plate with no \cn\!\!. The plates were incubated at 25°C for 48 hours, after which we observed an increase in the spread-area of the \pa colony. In figure \ref{fig:enhancedspreadofPAonCNlawn}b, we show the difference between the spread of \pa on half a lawn of \cn and on a half lawn without it. 
Figure \ref{fig:enhancedspreadofPAonCNlawn}c shows the dynamics of the spread of \pa (area covered and enhancement in spread), imaged every 6 hours. We found that \pa colony area increased rapidly after 12 hours of incubation with \cn lawn, where we recorded the maximum enhancement in the spread about ten times (see Movie S1). This was followed by a linear increase in area of \pa\!\!, which gave rise to a saturation in the enhancement, to about 3 times at 48 hour. Clearly, our study shows that the presence of \cn increases the spread of \pa\!\!.


\begin{figure}
\centering
\includegraphics[width=1\linewidth]{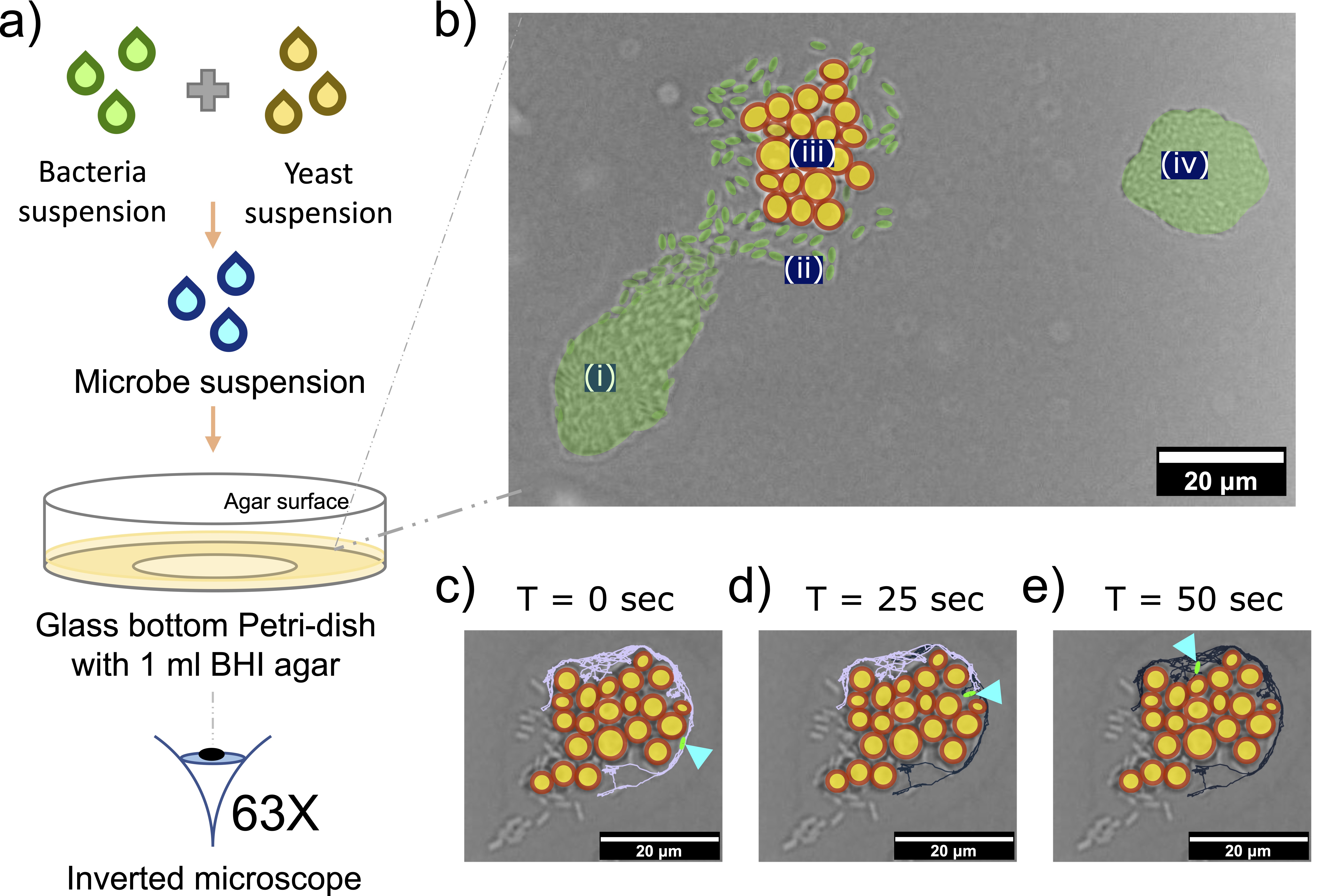}
\caption{\textbf{\pa exhibits exploratory behavior around \cn\!\!.} (a) Steps involved in preparing the \pa – \cn interaction assay for observation under an inverted microscope with extended working distance 63X objective lens focused on air-agar interface in glass bottom dish. The green drops represent bacterial suspension containing 0.2 $\mu$l of culture, the yellow drop represents yeast suspension containing 2  $\mu$l of culture, and the blue drop represents the mixture of the above. 
(b) Snapshot of the interacting microcolonies of \pa and \cn\!\!. The growing colony of \pa is marked as (i) and (iv) (pseudo-colored as green), whereas colony of \cn is marked as (iii) (pseudo-colored as yellow). The colony of \pa which exhibit exploratory motility is indicated as (ii). The colonies of \pa far from \cn do not exhibit this behavior: marked as (iv). 
(c)-(e) Snapshots of the microbes at different time intervals (of 25 s each). A single bacterium is marked with a cyan triangle. It explores the neighborhood of the \cn colony. The unexplored tracks around \cn colony are represented with white lines. The scale bar represents 20 $\mu$m.}
\label{fig:exploratorybehaviourPA}
\end{figure}

\subsection*{\pa cells exhibit exploratory behavior around \cn microcolonies}
To understand how the presence of \cn affects the spread of \pa\!\!, we examine the interactions between these cells at length and time scales corresponding to their growth and division. We used a long working distance 63X objective lens mounted to an inverted microscope to investigate the growth and movement of the bacterial and yeast cells in a 35 mm glass bottom petri dish with BHI agar at the air-agar interface. 
For this study, we mixed both cultures together and evenly spread them on the plate, allowing the \pa and \cn populations to form microcolonies on agar (Figure \ref{fig:exploratorybehaviourPA}a). 
Both the cells spread in the 2D agar surface as they grew and divided to accommodate new cells. We did not observe any active motility as long as the colonies of \pa and \cn were far away from each other.
However, when \pa cells came in proximity (say, around $10 \mu m$) to \cn microcolonies, they began to exhibit rapid movement. The newly growing \pa cells began to move towards the \cn microcolony, surrounding it, as shown in Figure \ref{fig:exploratorybehaviourPA}b (also see Movie S2). 
We term this behavior where \pa shows enhanced motility in the proximity of \cn as \textit{exploratory behavior}. 
Tracking the movement of individual cells in the field of view, we were able to compare the movements of these cells near the \cn microcolony and far away (regions marked \textit{i} or \textit{iv} and \textit{ii} in figure~\ref{fig:exploratorybehaviourPA}b). 
We find that in a $50 s$ duration, cells were able to explore the whole neighborhood of a small growing \cn microcolony (see tracked lines in Figure \ref{fig:exploratorybehaviourPA}c-e; also see Movie S3 a). We also found that the \pa cells at a distance of about $\sim 40 \mu m$ from \cn cells did not show appreciable movement. 

\begin{figure*}
\centering
\includegraphics[width=1\linewidth]{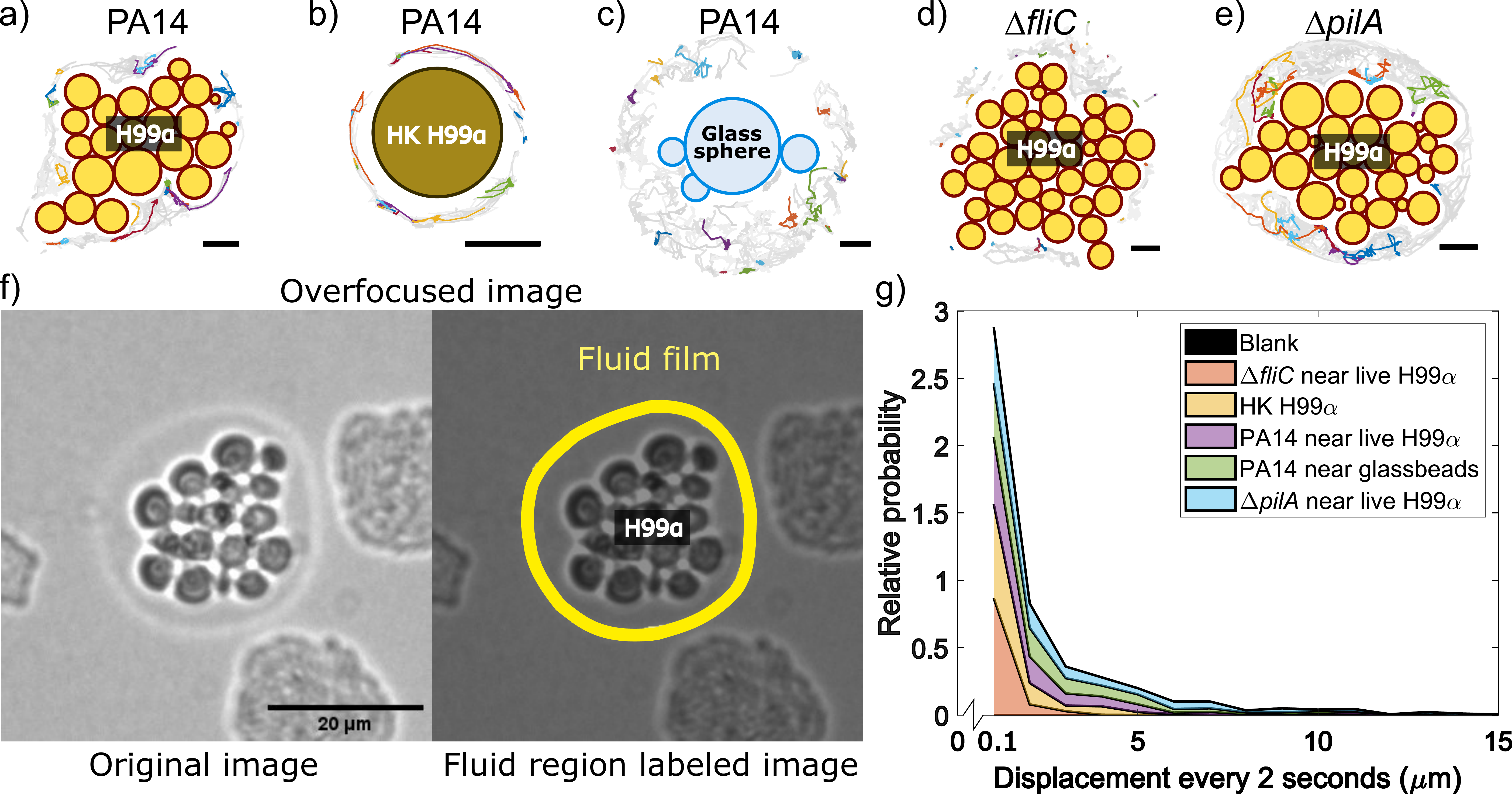}
\caption{\textbf{Exploratory behaviour of \pa is due to fluid film around \cn\!\!.} Tracks of \pa wild type PA14 strain of bacteria, over a period of two seconds, with twenty of them colored and the rest shown in gray: (a) around live \cn cells represented with yellow circles, (b) heat killed \cn cells, (c) inert glass spheres represented as light blue circles.
(d) Similarly, tracks of a flagellum defective $\Delta$\textit{fliC} variant of \pa around \cn\!\!.
(e) Similarly, tracks of a pilus defective $\Delta$\textit{pilA} variant of \pa around \cn\!\!.
(f) Original and over-focused snapshots of \cn micro-colony showing the presence of fluid film around the cells marked with a yellow circle. (g) Relative probability of bacterial cells producing a displacement $>0.1~\mu m$. The scale bar in (a - e) corresponds to $5~\mu m$.
}
\label{fig:fluidmediatedexploratorybehaviour}
\end{figure*}

\subsection*{Exploratory behavior is caused by the fluid layer accumulated around \cn\!\!} 

Next, we set out to identify what caused the observed exploratory behavior in \pa\!\!. The common understanding in the field would dictate that this phenomenon could arise as a chemotactic response of \pa to specific markers secreted by \cn \cite{Limoli2019,Rella2012}.
To test this hypothesis, we replaced the \cn cells with ones that were heat-killed (at $60 ^\circ C$ for one hour) and observed the motility of \pa around these cells under the microscope. 
Surprisingly, we found that \pa exhibited a qualitatively similar exploratory behavior just like it did with the live yeast cells (see Figure \ref{fig:fluidmediatedexploratorybehaviour}a and b; also see Movie S3 b). This shows that the observed exploratory behavior cannot be attributed to the active chemical secretion of the yeast cells. 
In addition, we employed inert glass spheres which were of the size  $5-15 \mu m$ (see SI figure S1) in the place of the heat-killed \cn cells to eliminate the possibility of any residual effects due to passive secretions from the dead cells. We found that \pa exhibited exploratory behavior even around the glass spheres (Figure \ref{fig:fluidmediatedexploratorybehaviour}c; also Movie S3 c).
These findings not only ruled out chemotaxis, the go-to explanation in the field, but also showed that the cause of the exploratory behavior must be physical in origin, based entirely on the morphological characteristics of \cn\!\!. 

Next, we wanted to understand what movement strategy the \pa cells used to achieve these larger displacements. We conducted the study using different strains of \pa with defects in the:
\begin{inlineenum}
    \item flagellum, $\Delta$\textit{fliC} (figure~\ref{fig:fluidmediatedexploratorybehaviour} d, Movie S3 d) and
    \item pilus (figure~\ref{fig:fluidmediatedexploratorybehaviour} e, Movie S3 e) $\Delta$\textit{pilA}.
\end{inlineenum}
We found that the $\Delta$\textit{fliC} variants were unable to produce the exploratory behavior observed in the wild type. $\Delta$\textit{pilA} variants did not show any decrease in motility (in fact, there was a slight increase). 
Figure~\ref{fig:fluidmediatedexploratorybehaviour} g shows the probability of displacements made by individual cells in $2 sec$.
To sum up, it is clear that the exploratory behavior exhibited by \pa is due to the flagellum they possess. 

Flagellum is used for swimming, and clearly, there is a need for a fluid layer around \cn if \pa cells had to swim.
In a recent study~\cite{Araujo2019}, Araujo and co-workers showed that water reservoirs are formed around graphite particles, which they externally introduced on the 0.5\% (w/v) agar plate. These particles were larger than the bacterial cells and allowed the flagellated bacteria to swim in this water layer. Interestingly, these graphite particles are similar in size to \cn cells. Nevertheless, the BHI agar media utilized in our study lacks any supplementary surfactant and consists of 99\% water. Further, the work by Xiao and Qian has shown that fluid can accumulate around inert micro-sized particles placed on a solid surface by capillary condensation as well \cite{Xiao2000}.  Hence, it could be the fluid accumulated around the \cn cells that causes the exploratory behavior. 
When we looked closely at our experiments, we found a boundary that showed the presence of fluid accumulation near the microcolonies of yeast cells (see figure~\ref{fig:fluidmediatedexploratorybehaviour} f). The Movie S4 clearly shows the enhancement in the motility of \pa once it reaches this fluid boundary. Hence, we conclude that the cause for the exploratory behavior is due to the swimming motility exhibited by \pa in the fluid accumulated around \cn microcolonies.

\begin{figure*}
\centering
\includegraphics[width=1\linewidth ]{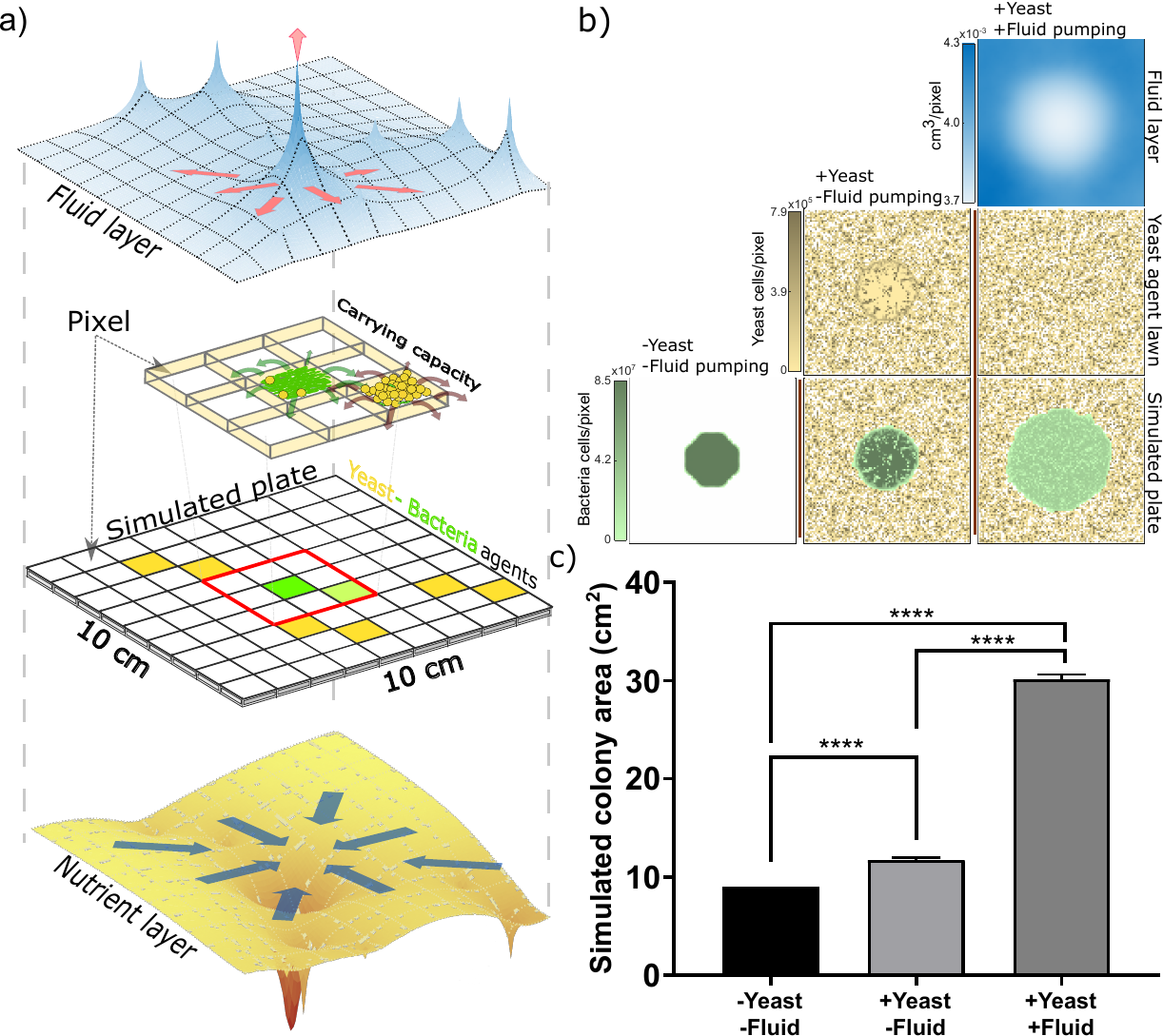}
\caption{\textbf{Model of the spread of \pa in a growing lawn of \cn}. (a) Illustration of the 2D model: the plate region is divided into pixels (100x100 units representing a physical area of 10 x 10 cm). Nutrient and the fluid layers: their formation, consumption and accumulation are independently modelled. Bacterial and Yeast cells in a pixel grow and divide to occupy the entire pixel (carrying capacity) before `spilling' over to neighbouring pixels.
(b) Simulation results at the end of 6 hours. Bacterial culture is seeded to the center-most pixel while the yeast culture is spread throughout the 2D domain, to mimic experiments as shown in~\ref{fig:enhancedspreadofPAonCNlawn} a. We show the fluid layer accumulated, the yeast colony occupancy and the spread of bacterial cells.
The first column corresponds to the case where there is no yeast lawn. In the second, the yeast lawn is present but the bacterial cells do not exhibit exploratory motility. Third, corresponds to the case we see in experiments where the lawn dynamically grows and the bacterial cells exhibit exploratory motility. 
(c) Area of spread of bacteria averaged over 100 independent realisations for three cases as elucidated above. Two-way ANOVA was used as a statistical test. The p values are depicted by: * p$<$0.05; ** p$<$0.01; *** p$<$0.001; **** p$<$0.0001.}
\label{fig:spreadingmodel_results}
\end{figure*}

\subsection*{Fluid layer mediated exploration enhances spreading}

Now the next step in our analysis is to test if the exploratory behavior, observed at the scale of the micro-organisms, is the primary mechanism for the enhanced migration of \pa cells at the scale of the plate. 
To this end, we construct a two-dimensional (2D) spatially explicit population model to simulate the dynamically changing landscape due to the growing yeast colonies and the resultant spreading of bacteria. The model incorporates the effect of exploratory behavior on the spreading of bacterial cells in the proximity of yeast colonies.

\paragraph*{Model.}
A 2D square domain is considered, that is comparable to the physical area of the agar plate. It is discretized into $m\times m$ smaller square regions, which we refer to as pixels (see Figure \ref{fig:spreadingmodel_results}a).
Cells are `placed' in these pixels: 
\begin{inlineenum}
    \item Yeast cells are distributed evenly across the 2D domain and,
    \item the bacterial cells are placed in the center of the 2D domain,
\end{inlineenum}
to reflect the initial conditions of the plate experiments (as shown in figure~\ref{fig:enhancedspreadofPAonCNlawn} a).
Cells grow and divide to occupy an entire pixel area.
The total number of cells that a pixel can accommodate is referred to as the carrying capacity; this depends on the area occupied by individual cells.
As cells grow in number, they consume the nutrient available in the pixel region and nutrients from the neighboring pixels diffuse to this pixel.
The yeast cells also pump water into the system through osmosis, causing the formation of a fluid layer around the proliferating yeast cells in the pixel. If more water is produced in one pixel it diffuses to its neighboring pixels.
When the cells divide and grow in excess, beyond the carrying capacity, the excess cells move to occupy the neighboring pixels: with more preference for nutrient-rich and water-rich pixels.

When the bacterial cells exhibit exploratory motility the effective area each cell occupies is larger than when it is immotile. This can be clearly seen in the regions marked as \textit{i} and \textit{ii} in figure~\ref{fig:exploratorybehaviourPA} b. Hence, when bacterial cells move into pixels containing water they fill up the pixel faster, resulting in a quick overflow of bacterial cells to the neighbouring pixels. However, this effect is only temporary. As the cells crowd in a pixel, larger densities constrain the flagellated motility of the bacterial cells, packing them more closely and restoring the effective area to the original area of the cells. More details of the model and implementation can be found in the Materials and methods section.

After seeding the same amount of \pa and \cn cells as in the experiments, we simulate the growth and spread of these cells till they cover the entire 2D domain. 
Our simulations show that the increased motility due to the exploratory behavior, which only temporarily influences the dynamics at the scale of a pixel, is sufficient to produce the spreading of \pa at the scale of the plate, as observed in the experiments (Figure \ref{fig:spreadingmodel_results} b third column; also see Movie S5).
We confirm this by comparing our results with two other simulations where:
\begin{inlineenum}
    \item the \cn lawn is absent and,
    \item the \cn lawn is present but \pa cells do not show any change in motility around \cn\!\!.
\end{inlineenum}
When the \cn lawn is not present, the \pa culture spreads radially in a rather uniform fashion, filling up all the pixels to its carrying capacity as shown in the first column of figure \ref{fig:spreadingmodel_results} b.
When \cn is present, its mere presence could increase the spread of \pa\!\!, since yeast takes up some of the space available. However, we find that this increase in spread is very small and is not comparable to the enhancement observed in experiments (see second column of figure \ref{fig:spreadingmodel_results} b). Interestingly, this scenario corresponds to the case when the flagella of \pa is defective ($\Delta$\textit{fliC}). When we plate $\Delta$\textit{fliC} in a lawn of \cn\!\!, we observe a very similar spread as predicted in our simulations (see SI figure S2).

\subsection*{Dynamically changing landscape regulates the spread of \pa}
While the exploratory motility of \pa around a \cn micro-colony enhances the spread of the bacterial cells locally, the spread observed at the level of the plate is a consequence of the dynamically changing fluid-landscape due to the growing \cn population in the neighbourhood.
Yeast cells not only increase the local spreading of the bacterial cells through the accumulated fluid layer, but also offer competition for space as they grow to occupy the area. 
Hence, one could expect the growth rates of the yeast lawn and the initial loading of the cells in the plate to affect the dynamics of the fluid landscape which in turn will influence the spread of bacteria (see SI figure S5, S6).

\begin{figure*}
\centering
\includegraphics[width=0.8\linewidth]{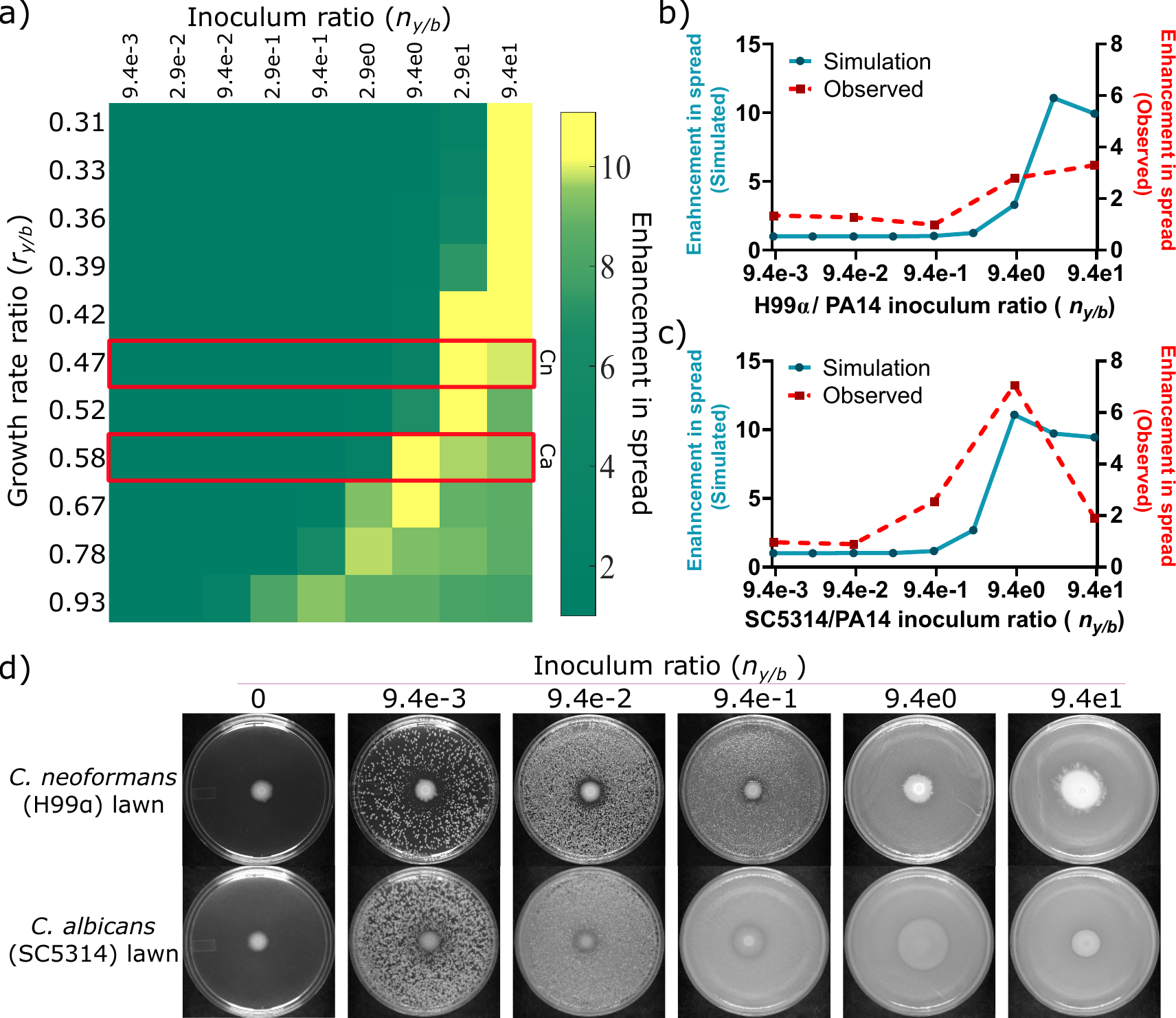}
\caption{\textbf{Spread of bacteria is mediated by the dynamics of the landscape}. 
(a) Heat map of the enhancement in the spread of bacteria as a function of the growth rate ratio $(r_{y/b})$ and inoculum ratio $(n_{y/b})$. The results are averaged over 100 independent simulations for every set of parameters.
(b, c) Comparing the experimentally observed spread of \pa on yeast lawn (\cn - b; \ca - c) for different dilutions $(n_{y/b})$ with predictions from the simulations. (d) Snapshots of the plate assay for different dilutions of \cn (top) and \ca (bottom) lawns.}
\label{fig:predictions_spreadingmodel}
\end{figure*}

To study these competing effects systematically, we carry out a number of simulations where we vary two key parameters in our model that affect the phenomenon:
\begin{inlineenum}
    \item the ratio of the \textit{number} of seeding cells $n_{y/b}$ and,
    \item the \textit{growth rates} of yeast and bacteria  cells $r_{y/b}$ (see figure~\ref{fig:predictions_spreadingmodel}a.
\end{inlineenum}
These parameters are also experimentally relevant because: we can change the initial number of cells added to the plate (dilutions) according to $n_{y/b}$; and since the exploratory phenomenon is not limited to the \cn cells, one can choose other kinds of yeast cells according to $r_{y/b}$.

We find that, in general, the enhancement in the spreading decreases when $n_{y/b}$ is very low. This corresponds to the case where the yeast colonies are scattered and not in proximity to the growing bacteria and hence, are unable to mediate their exploratory behavior.
However, with increasing $r_{y/b}$, we also find the emergence of an optimum number of yeast cells, with respect to $n_{y/b}$, that achieve the maximum enhancement in the spread of bacterial cells.
When $n_{y/b}$ is very high, it gives rise to competition for space between the growing microbe populations resulting in a reduced spread of the motile bacteria.
When the rate of growth of yeast cells $r_{y/b}$ is higher, the number of cells needed to mediate the optimal spread becomes lower.

To test the validity of our predictions, we carried out two sets of experiments. The spread of \pa was tested:
\begin{inlineenum}
    \item for a range of $n_{y/b}$ by changing the number of yeast added to the culture and,
    \item with two different yeast cells, \cn and \ca\!\!, where the latter grows $1.2$ times faster than the former.
\end{inlineenum}
The experimental observations of the enhancement in spread show a good qualitative match with our model predictions in identifying the conditions for optimal spread.
The enhancement in spread observed in the presence of \cn increases with $n_{y/b}$; the optimum lies close to the maximum value of $n_{y/b}$ tested, as shown in figure~\ref{fig:predictions_spreadingmodel} b. At the same time, since the dynamics of \ca are faster, we find that the optimum spreading occurs for a lower value of $n_{y/b}$ as seen in figure~\ref{fig:predictions_spreadingmodel} c.

In addition, we observe that the low values of enhancement observed when $n_{y/b}$ values are low, correspond with a `patchy' yeast lawn (see figure~\ref{fig:predictions_spreadingmodel} d). This finding corroborates with our exploratory behavior mediated enhancement in the spread, which requires proximity to the growing yeast lawn.
We also see that the slowly growing \cn lawn becomes patchy for a lower inoculum ratio in comparison to the faster growing \ca lawn, which results in the shift in the optimum loading observed with increasing $r_{y/b}$, as shown in figure~\ref{fig:predictions_spreadingmodel} a-c.

\subsection*{Final remarks}
In conclusion, our findings support the existence of a physical basis for the interactions occurring between motile bacteria and their immotile neighbors. The findings of our study indicate that alterations in the physical features of growth environments, such as the presence of accumulated fluid layers, might exert an influence on the dissemination and growth attributes of some microorganisms, thereby significantly affecting the composition of microhabitats. This phenomenon may elucidate alterations in microhabitats observed in natural environments, as well as pathophysiological situations impacting animal well-being.

\section*{Materials and methods}
\subsection*{Microbes and growth conditions}
\pafull (PA14 WT), PA14\textit{$\Delta$pilA}, PA14\textit{$\Delta$fliC},  \cn (H99$\alpha$), and \ca (SC5314) were used in this study. The WT PA14, \textit{$\Delta$pilA}, and \textit{$\Delta$fliC} are routinely cultured (at 37°C) in the Luria-Bertani (LB) media (HiMedia\textregistered, M575). Additional media used in this study includes the Brain Heart Infusion (BHI) medium (HiMedia\textregistered, M210) with 1\% Bacto agar (BD Difco\textsuperscript{TM}, 214010) and, the Yeast Extract-Peptone-Dextrose (YPD) medium (HiMedia\textregistered, M1363). The H99$\alpha$ and SC5314 were cultured in YPD broth and incubated at 25°C on a rotor moving at 25 RPM. 

\subsection*{Interaction assay}
All interaction assays between \pa strains and fungal pathogens were performed on BHI agar incubated at 25°C.
We prepared and autoclaved media containing 3.7\% w/v BHI media and 1\% w/v Bacto agar. This media was used in plate assay as well as plate-based microscopy assay.

\subsubsection*{Plate assay}
We poured 25 ml of BHI agar media into a 90 mm petri dish. \cn H99$\alpha$ or \ca culture SC5314 culture was grown in YPD for 12 hours ($OD_{600}$ $\sim$3, $\sim$2.25e7 cells). 0.5 ml of H99$\alpha$ culture was poured on a BHI agar plate, swirled, and allowed to dry for 50 minutes at room temperature (RT). \pa PA14 culture ($OD_{600}$ $\sim$1.5, $\sim$2.4e6 cells), grown in LB media for 12 hours. 2 $\mu$l of PA14 culture was spotted at the center of the BHI agar plate with and without H99$\alpha$ lawn and allowed to dry at RT for 20 minutes. All inoculated plates were incubated at 25°C for 48 hours.

\subsubsection*{Plate-based microscopy assay}
1 ml of BHI agar media was spread in a 35 mm glass bottom petri dish (ibidi\textregistered, 81218) and allowed to solidify. An inoculation mix was prepared by adding 1 $\mu$l of \cn culture, or heat-killed \cn\!\!, or 0.01\% (w/v) glass spheres (Sigma-Aldrich, 440345) in PBS solution 200 $\mu$l of PBS  solution. To this 0.2 $\mu$l of \pa culture was added. The solution was added to the surface of the agar and swirled. Any remaining liquid was gently removed using a pipette. The plates were kept for drying at RT for 30 minutes before performing microscopy.

Interaction between the cells (microcolonies) on the agar surface of the glass bottom Petri plate was imaged using 63X, a long working distance, dry objective lens (Leica\textsuperscript{TM}, 11506216) attached to Leica DMi8 inverted microscope (Figure \ref{fig:exploratorybehaviourPA}a). 

\subsection*{Image segmentation and tracking}
We exported the image from Leica LAS X and performed spectral filtering using MATLAB. We then used the Trackmate plugin of ImageJ to track the individual bacteria and generate the coordinate table \cite{Ershov2022}. This table was processed in MATLAB to generate the displacement histograms. All MATLAB codes used were custom-written.

\subsection*{Parameter extraction from experimental images}
We recorded the growth and division of isolated \pa\!\!, \cn, and \ca cells on BHI agar in a 35 mm glass bottom petri dish and captured time-lapse images with an interval of 30 sec and 34 ms. We computed the growth rates of these cells from the time taken for doubling. We estimated the average growth rates by repeating the process over ten independent sets of cells from three different time-lapse videos captured on different days. Then, using ImageJ, we estimated the area of \pa by measuring its width and length to account for its non-circular spherocylindrical shape. For \cn\!\!, we estimated the area by measuring its diameter. The estimates were averaged over ten independent measurements, taken from three images captured on different days.

\subsection*{2D spread and growth model}
A 2D grid, representing $10$ x $10$ cm of the agar plate, is divided into 100 x 100 pixels. We evenly distributed the yeast agents (\textit{N$_{yeast}$}) on this grid while bacteria (\textit{N$_{bacteria}$}) were kept in the center, such that there exist 9.375 yeast cells per bacteria agent. We allowed these agents to grow and divide. The bacteria agents divide at the growth rate of \textit{P. aeruginosa} i.e., 1.71 cells/hr, whereas yeast agents divide at the growth rate of \textit{C. neoformans} i.e., 0.8 cells/hr. We know that the area of a pixel is 0.01 cm\textsuperscript{2}; therefore, a single pixel can hold a limited number of agents; we termed it as the carrying capacity of a pixel. We estimated the area occupied by a bacterium agent by assuming it to be a chain of three circles with a radius of  0.5 $\mu$m, which represents the spherocylindrical shape of \textit{P. aeruginosa} with 1.5 $\mu$m of length and 0.5 $\mu$m of width. Similarly, we estimated the area occupied by each yeast agent by considering a yeast to be circular with a radius of 4.5 $\mu$m. 
Each \pa cell is assumed to occupy $a_p = 5.9\times 10^{-9}~cm^2$ while a \cn cell occupies an area of $a_y = 6.4\times 10^{-7}~cm^2$. 

We considered the presence of 10 M nutrient (\textit{C}) for consumption by the individual agents at the rate mentioned in Table \ref{table:1} to fuel their division and growth. As the nutrient depletes in occupied pixels, the nutrient from neighboring pixels will diffuse into it following the diffusion equation \ref{eq:1} with diffusivity mentioned in Table \ref{table:1}, replenishing the consumed nutrient. 
 \begin{equation} \label{eq:1}
 \begin{split}
 \frac{\partial C} {\partial t} & = D_{Nutrient}\ \frac{\partial^2\ C(x,t)} {\partial x^2} -(X_{Bacteria} * N_{Bacteria}(x)) \\ & -(X_{Yeast})* N_{yeast}(x))
 \end{split}
 \end{equation}
Further, each growing yeast agent will pump fluid from the agar surface (\textit{W}) at the rate mentioned in Table \ref{table:1}. The fluid will diffuse into the neighboring pixels following the equation \ref{eq:2} with parameters mentioned in Table \ref{eq:1}, giving rise to a fluid layer. 
 \begin{equation} \label{eq:2}
 \frac{\partial W} {\partial t} = D_{Fluid}\ \frac{\partial^2\ W(x,t)} {\partial x^2} +(P_{Fluid} * N_{yeast}(x)) 
 \end{equation}
 
When the dividing agents exceed the carrying capacity of a pixel, \textit{i.e.,} $n_p\times a_p + n_y\times a_y > 100/m^2$, it results in overflow into its neighborhood. Here, $n_p$ and $n_y$ are the numbers of bacteria and yeast cells, respectively. How the overflow is distributed in the neighboring pixels depends on the nutrient and fluid presence (only for bacteria spread) in those neighboring pixels. 
We assumed that the yeast agents could passively pump fluid from the substrate to form a fluid pool around themselves at the rate mentioned in Table \ref{table:1}. This fluid film diffuses into the neighborhood following the diffusion equation \ref{eq:2} with diffusivity mentioned in Table \ref{table:1}. On the other hand, we realize that the constant pumping of fluid would make it possible for bacterium agents to move through space in a three-dimensional manner. As a result, in order to restrict the scope of our model to only two dimensions, we made the assumption that the pumping of fluid would cease in the vicinity of yeast cells that are encompassed by bacterial agents.

We incorporated the exploratory behavior of bacteria by considering the change in the effective area $a_p$ of a bacterium in relation to the presence of the fluid. We assumed that the bacteria agents, when first encountering the fluid film, would exhibit exploratory behavior and thus display an increased effective area: \textit{i.e.,} $a_p = a_p^0 (k+w^2)/k$ where k is a constant termed as `fluid effect',  $a_p^0$ is the area of an immotile cell.
However, the effective area will soon be restored ($a_p$ to $a_p^0$) as the number of bacteria increases, causing the jamming of bacteria in the fluid pool. At this time we also enforced that only those pixels containing only yeast cells would actively pump fluid and contribute to the fluid pool in the plate.
Hence, the pixels containing both yeast and bacteria agents would not add to the fluid pool. Therefore, the leading edge of the spreading bacterial colony will experience higher fluid accumulation that aids swimming and further expansion, whereas the bacteria within the colony will fill the available region densely.

 \begin{table}
 \centering
 \caption{Parameter used in bacteria- yeast interaction model}
 \begin{tabular}{||c c c||} 
  \hline
 Parameters & Values & Ref \\ [0.5ex] 
  \hline\hline
 Nutrient consumption rate (X$_{Yeast}$) & 16e-9 mol/kg/cell/hr &  \cite{Otterstedt2004}\\
 Nutrient consumption rate (X$_{Bacteria}$) & 7e-9 mol/kg/cell/hr &  \cite{Fischer2004}\\
 Fluid pump rate (P$_{fluid}$) & 0.3e-9 cm\textsuperscript{3}/cell/hr &  \cite{Armstrong2015}\\
 D$_{Nutrient}$ & 0.0446 cm\textsuperscript{2}/hr & \cite{Arnikar1980} \\
 D$_{Fluid}$  & 0.0828 cm\textsuperscript{2}/hr & \cite{Holz2000} \\[1ex] 
 \hline
 \end{tabular}
 \label{table:1}
 \end{table}

\subsection*{Data and codes}
All the relevant data and the codes can be found in: \url{https://zenodo.org/record/8330043}

\section*{Acknowledgements}
We would like to thank Prof. Frederick
Ausubel for providing the PA14 strain, Prof. Zemer Gitai for providing $\Delta$\textit{flic} and $\Delta$\textit{pilA} strain of \textit{Pseudomonas aeruginosa}. We would also like to thank Prof. Kaustuv Sanyal for providing the H99$\alpha$ strain of \textit{Cryptococcus neoformans} and the SC5314 strain of \textit{Candida albicans}. VS thanks the senior fellowship (IA/S/21/1/505655) from the Wellcome Trust/DBT India Alliance for funding. AK thanks SERB grant (CRG/2022/005381) for the funding. DRM thanks the DST INSPIRE faculty award (DST/INSPIRE/04/
2017/002985) for funding.

\bibliography{references.bib}

\begin{thebibliography}{39}%
\makeatletter
\providecommand \@ifxundefined [1]{%
 \@ifx{#1\undefined}
}%
\providecommand \@ifnum [1]{%
 \ifnum #1\expandafter \@firstoftwo
 \else \expandafter \@secondoftwo
 \fi
}%
\providecommand \@ifx [1]{%
 \ifx #1\expandafter \@firstoftwo
 \else \expandafter \@secondoftwo
 \fi
}%
\providecommand \natexlab [1]{#1}%
\providecommand \enquote  [1]{``#1''}%
\providecommand \bibnamefont  [1]{#1}%
\providecommand \bibfnamefont [1]{#1}%
\providecommand \citenamefont [1]{#1}%
\providecommand \href@noop [0]{\@secondoftwo}%
\providecommand \href [0]{\begingroup \@sanitize@url \@href}%
\providecommand \@href[1]{\@@startlink{#1}\@@href}%
\providecommand \@@href[1]{\endgroup#1\@@endlink}%
\providecommand \@sanitize@url [0]{\catcode `\\12\catcode `\$12\catcode
  `\&12\catcode `\#12\catcode `\^12\catcode `\_12\catcode `\%12\relax}%
\providecommand \@@startlink[1]{}%
\providecommand \@@endlink[0]{}%
\providecommand \url  [0]{\begingroup\@sanitize@url \@url }%
\providecommand \@url [1]{\endgroup\@href {#1}{\urlprefix }}%
\providecommand \urlprefix  [0]{URL }%
\providecommand \Eprint [0]{\href }%
\providecommand \doibase [0]{https://doi.org/}%
\providecommand \selectlanguage [0]{\@gobble}%
\providecommand \bibinfo  [0]{\@secondoftwo}%
\providecommand \bibfield  [0]{\@secondoftwo}%
\providecommand \translation [1]{[#1]}%
\providecommand \BibitemOpen [0]{}%
\providecommand \bibitemStop [0]{}%
\providecommand \bibitemNoStop [0]{.\EOS\space}%
\providecommand \EOS [0]{\spacefactor3000\relax}%
\providecommand \BibitemShut  [1]{\csname bibitem#1\endcsname}%
\let\auto@bib@innerbib\@empty
\bibitem [{\citenamefont {Araujo}\ \emph {et~al.}(2019)\citenamefont {Araujo},
  \citenamefont {Chen}, \citenamefont {Mani},\ and\ \citenamefont
  {Tang}}]{Araujo2019}%
  \BibitemOpen
  \bibfield  {author} {\bibinfo {author} {\bibnamefont {Araujo}, \bibfnamefont
  {G.}}, \bibinfo {author} {\bibnamefont {Chen}, \bibfnamefont {W.}}, \bibinfo
  {author} {\bibnamefont {Mani}, \bibfnamefont {S.}}, and\ \bibinfo {author}
  {\bibnamefont {Tang}, \bibfnamefont {J.~X.}},\ }\bibfield  {title} {\enquote
  {\bibinfo {title} {Orbiting of flagellated bacteria within a thin fluid film
  around micrometer-sized particles},}\ }\href
  {https://doi.org/10.1016/j.bpj.2019.06.005} {\bibfield  {journal} {\bibinfo
  {journal} {Biophysical Journal}\ }\textbf {\bibinfo {volume} {117}},\
  \bibinfo {pages} {346--354} (\bibinfo {year} {2019})},\ \bibinfo {note} {doi:
  10.1016/j.bpj.2019.06.005}\BibitemShut {NoStop}%
\bibitem [{\citenamefont {Armstrong}\ \emph {et~al.}(2015)\citenamefont
  {Armstrong}, \citenamefont {Evseev}, \citenamefont {Koroteev},\ and\
  \citenamefont {Berg}}]{Armstrong2015}%
  \BibitemOpen
  \bibfield  {author} {\bibinfo {author} {\bibnamefont {Armstrong},
  \bibfnamefont {R.~T.}}, \bibinfo {author} {\bibnamefont {Evseev},
  \bibfnamefont {N.}}, \bibinfo {author} {\bibnamefont {Koroteev},
  \bibfnamefont {D.}}, and\ \bibinfo {author} {\bibnamefont {Berg},
  \bibfnamefont {S.}},\ }\bibfield  {title} {\enquote {\bibinfo {title}
  {Modeling the velocity field during haines jumps in porous media},}\ }\href
  {https://doi.org/10.1016/j.advwatres.2015.01.008} {\bibfield  {journal}
  {\bibinfo  {journal} {Advances in Water Resources}\ }\textbf {\bibinfo
  {volume} {77}},\ \bibinfo {pages} {57--68} (\bibinfo {year}
  {2015})}\BibitemShut {NoStop}%
\bibitem [{\citenamefont {Arnikar}\ \emph {et~al.}(1980)\citenamefont
  {Arnikar}, \citenamefont {Patil}, \citenamefont {Adhyapak},\ and\
  \citenamefont {Potdar}}]{Arnikar1980}%
  \BibitemOpen
  \bibfield  {author} {\bibinfo {author} {\bibnamefont {Arnikar}, \bibfnamefont
  {H.~J.}}, \bibinfo {author} {\bibnamefont {Patil}, \bibfnamefont {S.~F.}},
  \bibinfo {author} {\bibnamefont {Adhyapak}, \bibfnamefont {N.~G.}}, and\
  \bibinfo {author} {\bibnamefont {Potdar}, \bibfnamefont {J.~K.}},\ }\bibfield
   {title} {\enquote {\bibinfo {title} {Diffusion of potassium chromate in agar
  gel},}\ }\href {https://doi.org/10.1524/zpch.1980.120.1.051} {\bibfield
  {journal} {\bibinfo  {journal} {Zeitschrift fur Physikalische Chemie}\
  }\textbf {\bibinfo {volume} {120}},\ \bibinfo {pages} {51--57} (\bibinfo
  {year} {1980})}\BibitemShut {NoStop}%
\bibitem [{\citenamefont {Belas}(2013)}]{Belas2013}%
  \BibitemOpen
  \bibfield  {author} {\bibinfo {author} {\bibnamefont {Belas}, \bibfnamefont
  {R.}},\ }\bibfield  {title} {\enquote {\bibinfo {title} {When the swimming
  gets tough, the tough form a biofilm},}\ }\href
  {https://doi.org/https://doi.org/10.1111/mmi.12354} {\bibfield  {journal}
  {\bibinfo  {journal} {Molecular Microbiology}\ }\textbf {\bibinfo {volume}
  {90}},\ \bibinfo {pages} {1--5} (\bibinfo {year} {2013})},\ \Eprint
  {https://arxiv.org/abs/https://onlinelibrary.wiley.com/doi/pdf/10.1111/mmi.12354}
  {https://onlinelibrary.wiley.com/doi/pdf/10.1111/mmi.12354} \BibitemShut
  {NoStop}%
\bibitem [{\citenamefont {Bubendorfer}\ \emph {et~al.}(2014)\citenamefont
  {Bubendorfer}, \citenamefont {Koltai}, \citenamefont {Rossmann},
  \citenamefont {Sourjik},\ and\ \citenamefont {Thormann}}]{Bubendorfer2014}%
  \BibitemOpen
  \bibfield  {author} {\bibinfo {author} {\bibnamefont {Bubendorfer},
  \bibfnamefont {S.}}, \bibinfo {author} {\bibnamefont {Koltai}, \bibfnamefont
  {M.}}, \bibinfo {author} {\bibnamefont {Rossmann}, \bibfnamefont {F.}},
  \bibinfo {author} {\bibnamefont {Sourjik}, \bibfnamefont {V.}}, and\ \bibinfo
  {author} {\bibnamefont {Thormann}, \bibfnamefont {K.~M.}},\ }\bibfield
  {title} {\enquote {\bibinfo {title} {Secondary bacterial flagellar system
  improves bacterial spreading by increasing the directional persistence of
  swimming},}\ }\href {https://doi.org/10.1073/pnas.1405820111} {\bibfield
  {journal} {\bibinfo  {journal} {Proceedings of the National Academy of
  Sciences}\ }\textbf {\bibinfo {volume} {111}},\ \bibinfo {pages}
  {11485--11490} (\bibinfo {year} {2014})},\ \Eprint
  {https://arxiv.org/abs/https://www.pnas.org/doi/pdf/10.1073/pnas.1405820111}
  {https://www.pnas.org/doi/pdf/10.1073/pnas.1405820111} \BibitemShut {NoStop}%
\bibitem [{\citenamefont {Conrad}\ \emph {et~al.}(2011)\citenamefont {Conrad},
  \citenamefont {Gibiansky}, \citenamefont {Jin}, \citenamefont {Gordon},
  \citenamefont {Motto}, \citenamefont {Mathewson}, \citenamefont {Stopka},
  \citenamefont {Zelasko}, \citenamefont {Shrout},\ and\ \citenamefont
  {Wong}}]{Conrad2011}%
  \BibitemOpen
  \bibfield  {author} {\bibinfo {author} {\bibnamefont {Conrad}, \bibfnamefont
  {J.}}, \bibinfo {author} {\bibnamefont {Gibiansky}, \bibfnamefont {M.}},
  \bibinfo {author} {\bibnamefont {Jin}, \bibfnamefont {F.}}, \bibinfo {author}
  {\bibnamefont {Gordon}, \bibfnamefont {V.}}, \bibinfo {author} {\bibnamefont
  {Motto}, \bibfnamefont {D.}}, \bibinfo {author} {\bibnamefont {Mathewson},
  \bibfnamefont {M.}}, \bibinfo {author} {\bibnamefont {Stopka}, \bibfnamefont
  {W.}}, \bibinfo {author} {\bibnamefont {Zelasko}, \bibfnamefont {D.}},
  \bibinfo {author} {\bibnamefont {Shrout}, \bibfnamefont {J.}}, and\ \bibinfo
  {author} {\bibnamefont {Wong}, \bibfnamefont {G.}},\ }\bibfield  {title}
  {\enquote {\bibinfo {title} {Flagella and pili-mediated near-surface
  single-cell motility mechanisms in \textit{P. aeruginosa}},}\ }\href
  {https://doi.org/https://doi.org/10.1016/j.bpj.2011.02.020} {\bibfield
  {journal} {\bibinfo  {journal} {Biophysical Journal}\ }\textbf {\bibinfo
  {volume} {100}},\ \bibinfo {pages} {1608--1616} (\bibinfo {year}
  {2011})}\BibitemShut {NoStop}%
\bibitem [{\citenamefont {Deveau}\ \emph {et~al.}(2018)\citenamefont {Deveau},
  \citenamefont {Bonito}, \citenamefont {Uehling}, \citenamefont {Paoletti},
  \citenamefont {Becker}, \citenamefont {Bindschedler}, \citenamefont
  {Hacquard}, \citenamefont {Hervé}, \citenamefont {Labbé}, \citenamefont
  {Lastovetsky}, \citenamefont {Mieszkin}, \citenamefont {Millet},
  \citenamefont {Vajna}, \citenamefont {Junier}, \citenamefont {Bonfante},
  \citenamefont {Krom}, \citenamefont {Olsson}, \citenamefont {van Elsas},\
  and\ \citenamefont {Wick}}]{Deveau2018}%
  \BibitemOpen
  \bibfield  {author} {\bibinfo {author} {\bibnamefont {Deveau}, \bibfnamefont
  {A.}}, \bibinfo {author} {\bibnamefont {Bonito}, \bibfnamefont {G.}},
  \bibinfo {author} {\bibnamefont {Uehling}, \bibfnamefont {J.}}, \bibinfo
  {author} {\bibnamefont {Paoletti}, \bibfnamefont {M.}}, \bibinfo {author}
  {\bibnamefont {Becker}, \bibfnamefont {M.}}, \bibinfo {author} {\bibnamefont
  {Bindschedler}, \bibfnamefont {S.}}, \bibinfo {author} {\bibnamefont
  {Hacquard}, \bibfnamefont {S.}}, \bibinfo {author} {\bibnamefont {Hervé},
  \bibfnamefont {V.}}, \bibinfo {author} {\bibnamefont {Labbé}, \bibfnamefont
  {J.}}, \bibinfo {author} {\bibnamefont {Lastovetsky}, \bibfnamefont {O.~A.}},
  \bibinfo {author} {\bibnamefont {Mieszkin}, \bibfnamefont {S.}}, \bibinfo
  {author} {\bibnamefont {Millet}, \bibfnamefont {L.~J.}}, \bibinfo {author}
  {\bibnamefont {Vajna}, \bibfnamefont {B.}}, \bibinfo {author} {\bibnamefont
  {Junier}, \bibfnamefont {P.}}, \bibinfo {author} {\bibnamefont {Bonfante},
  \bibfnamefont {P.}}, \bibinfo {author} {\bibnamefont {Krom}, \bibfnamefont
  {B.~P.}}, \bibinfo {author} {\bibnamefont {Olsson}, \bibfnamefont {S.}},
  \bibinfo {author} {\bibnamefont {van Elsas}, \bibfnamefont {J.~D.}}, and\
  \bibinfo {author} {\bibnamefont {Wick}, \bibfnamefont {L.~Y.}},\ }\bibfield
  {title} {\enquote {\bibinfo {title} {{Bacterial–fungal interactions:
  Ecology, mechanisms and challenges}},}\ }\href
  {https://doi.org/10.1093/femsre/fuy008} {\bibfield  {journal} {\bibinfo
  {journal} {FEMS Microbiology Reviews}\ }\textbf {\bibinfo {volume} {42}},\
  \bibinfo {pages} {335--352} (\bibinfo {year} {2018})},\ \Eprint
  {https://arxiv.org/abs/https://academic.oup.com/femsre/article-pdf/42/3/335/25030303/fuy008.pdf}
  {https://academic.oup.com/femsre/article-pdf/42/3/335/25030303/fuy008.pdf}
  \BibitemShut {NoStop}%
\bibitem [{\citenamefont {Ershov}\ \emph {et~al.}(2022)\citenamefont {Ershov},
  \citenamefont {Phan}, \citenamefont {Pylvänäinen}, \citenamefont {Rigaud},
  \citenamefont {Blanc}, \citenamefont {Charles-Orszag}, \citenamefont
  {Conway}, \citenamefont {Laine}, \citenamefont {Roy}, \citenamefont
  {Bonazzi}, \citenamefont {Duménil}, \citenamefont {Jacquemet},\ and\
  \citenamefont {Tinevez}}]{Ershov2022}%
  \BibitemOpen
  \bibfield  {author} {\bibinfo {author} {\bibnamefont {Ershov}, \bibfnamefont
  {D.}}, \bibinfo {author} {\bibnamefont {Phan}, \bibfnamefont {M.-S.}},
  \bibinfo {author} {\bibnamefont {Pylvänäinen}, \bibfnamefont {J.~W.}},
  \bibinfo {author} {\bibnamefont {Rigaud}, \bibfnamefont {S.~U.}}, \bibinfo
  {author} {\bibnamefont {Blanc}, \bibfnamefont {L.~L.}}, \bibinfo {author}
  {\bibnamefont {Charles-Orszag}, \bibfnamefont {A.}}, \bibinfo {author}
  {\bibnamefont {Conway}, \bibfnamefont {J.~R.~W.}}, \bibinfo {author}
  {\bibnamefont {Laine}, \bibfnamefont {R.~F.}}, \bibinfo {author}
  {\bibnamefont {Roy}, \bibfnamefont {N.~H.}}, \bibinfo {author} {\bibnamefont
  {Bonazzi}, \bibfnamefont {D.}}, \bibinfo {author} {\bibnamefont {Duménil},
  \bibfnamefont {G.}}, \bibinfo {author} {\bibnamefont {Jacquemet},
  \bibfnamefont {G.}}, and\ \bibinfo {author} {\bibnamefont {Tinevez},
  \bibfnamefont {J.-Y.}},\ }\bibfield  {title} {\enquote {\bibinfo {title}
  {Trackmate 7: integrating state-of-the-art segmentation algorithms into
  tracking pipelines},}\ }\href {https://doi.org/10.1038/s41592-022-01507-1}
  {\bibfield  {journal} {\bibinfo  {journal} {Nature Methods}\ }\textbf
  {\bibinfo {volume} {19}},\ \bibinfo {pages} {829--832} (\bibinfo {year}
  {2022})}\BibitemShut {NoStop}%
\bibitem [{\citenamefont {Fischer}, \citenamefont {Zamboni},\ and\
  \citenamefont {Sauer}(2004)}]{Fischer2004}%
  \BibitemOpen
  \bibfield  {author} {\bibinfo {author} {\bibnamefont {Fischer}, \bibfnamefont
  {E.}}, \bibinfo {author} {\bibnamefont {Zamboni}, \bibfnamefont {N.}}, and\
  \bibinfo {author} {\bibnamefont {Sauer}, \bibfnamefont {U.}},\ }\bibfield
  {title} {\enquote {\bibinfo {title} {High-throughput metabolic flux analysis
  based on gas chromatography–mass spectrometry derived 13c constraints},}\
  }\href {https://doi.org/10.1016/j.ab.2003.10.036} {\bibfield  {journal}
  {\bibinfo  {journal} {Analytical Biochemistry}\ }\textbf {\bibinfo {volume}
  {325}},\ \bibinfo {pages} {308--316} (\bibinfo {year} {2004})}\BibitemShut
  {NoStop}%
\bibitem [{\citenamefont {Flemming}\ \emph {et~al.}(2016)\citenamefont
  {Flemming}, \citenamefont {Wingender}, \citenamefont {Szewzyk}, \citenamefont
  {Steinberg}, \citenamefont {Rice},\ and\ \citenamefont
  {Kjelleberg}}]{Flemming2016}%
  \BibitemOpen
  \bibfield  {author} {\bibinfo {author} {\bibnamefont {Flemming},
  \bibfnamefont {H.-C.}}, \bibinfo {author} {\bibnamefont {Wingender},
  \bibfnamefont {J.}}, \bibinfo {author} {\bibnamefont {Szewzyk}, \bibfnamefont
  {U.}}, \bibinfo {author} {\bibnamefont {Steinberg}, \bibfnamefont {P.}},
  \bibinfo {author} {\bibnamefont {Rice}, \bibfnamefont {S.~A.}}, and\ \bibinfo
  {author} {\bibnamefont {Kjelleberg}, \bibfnamefont {S.}},\ }\bibfield
  {title} {\enquote {\bibinfo {title} {{Biofilms: An emergent form of bacterial
  life}},}\ }\href {https://doi.org/10.1038/nrmicro.2016.94} {\bibfield
  {journal} {\bibinfo  {journal} {Nature Reviews Microbiology}\ }\textbf
  {\bibinfo {volume} {14}},\ \bibinfo {pages} {563--575} (\bibinfo {year}
  {2016})}\BibitemShut {NoStop}%
\bibitem [{\citenamefont {Gode-Potratz}\ \emph {et~al.}(2011)\citenamefont
  {Gode-Potratz}, \citenamefont {Kustusch}, \citenamefont {Breheny},
  \citenamefont {Weiss},\ and\ \citenamefont {McCarter}}]{Potratz2011}%
  \BibitemOpen
  \bibfield  {author} {\bibinfo {author} {\bibnamefont {Gode-Potratz},
  \bibfnamefont {C.~J.}}, \bibinfo {author} {\bibnamefont {Kustusch},
  \bibfnamefont {R.~J.}}, \bibinfo {author} {\bibnamefont {Breheny},
  \bibfnamefont {P.~J.}}, \bibinfo {author} {\bibnamefont {Weiss},
  \bibfnamefont {D.~S.}}, and\ \bibinfo {author} {\bibnamefont {McCarter},
  \bibfnamefont {L.~L.}},\ }\bibfield  {title} {\enquote {\bibinfo {title}
  {Surface sensing in \textit{Vibrio parahaemolyticus} triggers a programme of
  gene expression that promotes colonization and virulence},}\ }\href
  {https://doi.org/https://doi.org/10.1111/j.1365-2958.2010.07445.x} {\bibfield
   {journal} {\bibinfo  {journal} {Molecular Microbiology}\ }\textbf {\bibinfo
  {volume} {79}},\ \bibinfo {pages} {240--263} (\bibinfo {year} {2011})},\
  \Eprint
  {https://arxiv.org/abs/https://onlinelibrary.wiley.com/doi/pdf/10.1111/j.1365-2958.2010.07445.x}
  {https://onlinelibrary.wiley.com/doi/pdf/10.1111/j.1365-2958.2010.07445.x}
  \BibitemShut {NoStop}%
\bibitem [{\citenamefont {Harshey}(2003)}]{Harshey2003}%
  \BibitemOpen
  \bibfield  {author} {\bibinfo {author} {\bibnamefont {Harshey}, \bibfnamefont
  {R.~M.}},\ }\bibfield  {title} {\enquote {\bibinfo {title} {Bacterial
  motility on a surface: Many ways to a common goal},}\ }\href
  {https://doi.org/10.1146/annurev.micro.57.030502.091014} {\bibfield
  {journal} {\bibinfo  {journal} {Annual Review of Microbiology}\ }\textbf
  {\bibinfo {volume} {57}},\ \bibinfo {pages} {249--273} (\bibinfo {year}
  {2003})},\ \bibinfo {note} {pMID: 14527279},\ \Eprint
  {https://arxiv.org/abs/https://doi.org/10.1146/annurev.micro.57.030502.091014}
  {https://doi.org/10.1146/annurev.micro.57.030502.091014} \BibitemShut
  {NoStop}%
\bibitem [{\citenamefont {Hershey}(2021)}]{HERSHEY2021}%
  \BibitemOpen
  \bibfield  {author} {\bibinfo {author} {\bibnamefont {Hershey}, \bibfnamefont
  {D.~M.}},\ }\bibfield  {title} {\enquote {\bibinfo {title} {Integrated
  control of surface adaptation by the bacterial flagellum},}\ }\href
  {https://doi.org/https://doi.org/10.1016/j.mib.2021.02.004} {\bibfield
  {journal} {\bibinfo  {journal} {Current Opinion in Microbiology}\ }\textbf
  {\bibinfo {volume} {61}},\ \bibinfo {pages} {1--7} (\bibinfo {year}
  {2021})}\BibitemShut {NoStop}%
\bibitem [{\citenamefont {Hibbing}\ \emph {et~al.}(2010)\citenamefont
  {Hibbing}, \citenamefont {Fuqua}, \citenamefont {Parsek},\ and\ \citenamefont
  {Peterson}}]{Hibbing2010}%
  \BibitemOpen
  \bibfield  {author} {\bibinfo {author} {\bibnamefont {Hibbing}, \bibfnamefont
  {M.~E.}}, \bibinfo {author} {\bibnamefont {Fuqua}, \bibfnamefont {C.}},
  \bibinfo {author} {\bibnamefont {Parsek}, \bibfnamefont {M.~R.}}, and\
  \bibinfo {author} {\bibnamefont {Peterson}, \bibfnamefont {S.~B.}},\
  }\bibfield  {title} {\enquote {\bibinfo {title} {{Bacterial competition:
  Surviving and thriving in the microbial jungle}},}\ }\href
  {https://doi.org/10.1038/nrmicro2259} {\bibfield  {journal} {\bibinfo
  {journal} {Nature Reviews Microbiology}\ }\textbf {\bibinfo {volume} {8}},\
  \bibinfo {pages} {15--25} (\bibinfo {year} {2010})}\BibitemShut {NoStop}%
\bibitem [{\citenamefont {Holz}, \citenamefont {Heil},\ and\ \citenamefont
  {Sacco}(2000)}]{Holz2000}%
  \BibitemOpen
  \bibfield  {author} {\bibinfo {author} {\bibnamefont {Holz}, \bibfnamefont
  {M.}}, \bibinfo {author} {\bibnamefont {Heil}, \bibfnamefont {S.~R.}}, and\
  \bibinfo {author} {\bibnamefont {Sacco}, \bibfnamefont {A.}},\ }\bibfield
  {title} {\enquote {\bibinfo {title} {Temperature-dependent self-diffusion
  coefficients of water and six selected molecular liquids for calibration in
  accurate 1h nmr pfg measurements},}\ }\href
  {https://doi.org/10.1039/b005319h} {\bibfield  {journal} {\bibinfo  {journal}
  {Physical Chemistry Chemical Physics}\ }\textbf {\bibinfo {volume} {2}},\
  \bibinfo {pages} {4740--4742} (\bibinfo {year} {2000})}\BibitemShut {NoStop}%
\bibitem [{\citenamefont {Kearns}(2010)}]{Kearns2010}%
  \BibitemOpen
  \bibfield  {author} {\bibinfo {author} {\bibnamefont {Kearns}, \bibfnamefont
  {D.~B.}},\ }\bibfield  {title} {\enquote {\bibinfo {title} {A field guide to
  bacterial swarming motility},}\ }\href {https://doi.org/10.1038/nrmicro2405}
  {\bibfield  {journal} {\bibinfo  {journal} {Nature Reviews Microbiology}\
  }\textbf {\bibinfo {volume} {8}},\ \bibinfo {pages} {634--644} (\bibinfo
  {year} {2010})}\BibitemShut {NoStop}%
\bibitem [{\citenamefont {Kelly}, \citenamefont {Dapsis},\ and\ \citenamefont
  {Lauffenburger}(1988)}]{Kelly1988}%
  \BibitemOpen
  \bibfield  {author} {\bibinfo {author} {\bibnamefont {Kelly}, \bibfnamefont
  {F.~X.}}, \bibinfo {author} {\bibnamefont {Dapsis}, \bibfnamefont {K.~J.}},
  and\ \bibinfo {author} {\bibnamefont {Lauffenburger}, \bibfnamefont
  {D.~A.}},\ }\bibfield  {title} {\enquote {\bibinfo {title} {{Effect of
  bacterial chemotaxis on dynamics of microbial competition}},}\ }\href
  {https://doi.org/10.1007/BF02018908} {\bibfield  {journal} {\bibinfo
  {journal} {Microbial Ecology}\ }\textbf {\bibinfo {volume} {16}},\ \bibinfo
  {pages} {115--131} (\bibinfo {year} {1988})}\BibitemShut {NoStop}%
\bibitem [{\citenamefont {Kerr}\ \emph {et~al.}(2002)\citenamefont {Kerr},
  \citenamefont {Riley}, \citenamefont {Feldman},\ and\ \citenamefont
  {Bohannan}}]{Kerr2002}%
  \BibitemOpen
  \bibfield  {author} {\bibinfo {author} {\bibnamefont {Kerr}, \bibfnamefont
  {B.}}, \bibinfo {author} {\bibnamefont {Riley}, \bibfnamefont {M.~A.}},
  \bibinfo {author} {\bibnamefont {Feldman}, \bibfnamefont {M.~W.}}, and\
  \bibinfo {author} {\bibnamefont {Bohannan}, \bibfnamefont {B.~J.~M.}},\
  }\bibfield  {title} {\enquote {\bibinfo {title} {{Local dispersal promotes
  biodiversity in a real-life game of rock–paper–scissors}},}\ }\href
  {https://doi.org/10.1038/nature00823} {\bibfield  {journal} {\bibinfo
  {journal} {Nature}\ }\textbf {\bibinfo {volume} {418}},\ \bibinfo {pages}
  {171--174} (\bibinfo {year} {2002})}\BibitemShut {NoStop}%
\bibitem [{\citenamefont {Kollaran}\ \emph {et~al.}(2019)\citenamefont
  {Kollaran}, \citenamefont {Joge}, \citenamefont {Kotian}, \citenamefont
  {Badal}, \citenamefont {Prakash}, \citenamefont {Mishra}, \citenamefont
  {Varma},\ and\ \citenamefont {Singh}}]{Kollaran2019}%
  \BibitemOpen
  \bibfield  {author} {\bibinfo {author} {\bibnamefont {Kollaran},
  \bibfnamefont {A.~M.}}, \bibinfo {author} {\bibnamefont {Joge}, \bibfnamefont
  {S.}}, \bibinfo {author} {\bibnamefont {Kotian}, \bibfnamefont {H.~S.}},
  \bibinfo {author} {\bibnamefont {Badal}, \bibfnamefont {D.}}, \bibinfo
  {author} {\bibnamefont {Prakash}, \bibfnamefont {D.}}, \bibinfo {author}
  {\bibnamefont {Mishra}, \bibfnamefont {A.}}, \bibinfo {author} {\bibnamefont
  {Varma}, \bibfnamefont {M.}}, and\ \bibinfo {author} {\bibnamefont {Singh},
  \bibfnamefont {V.}},\ }\bibfield  {title} {\enquote {\bibinfo {title}
  {Context-specific requirement of forty-four two-component loci in pseudomonas
  aeruginosa swarming},}\ }\href
  {https://doi.org/https://doi.org/10.1016/j.isci.2019.02.028} {\bibfield
  {journal} {\bibinfo  {journal} {iScience}\ }\textbf {\bibinfo {volume}
  {13}},\ \bibinfo {pages} {305--317} (\bibinfo {year} {2019})}\BibitemShut
  {NoStop}%
\bibitem [{\citenamefont {Konopka}(2009)}]{Konopka2009}%
  \BibitemOpen
  \bibfield  {author} {\bibinfo {author} {\bibnamefont {Konopka}, \bibfnamefont
  {A.}},\ }\bibfield  {title} {\enquote {\bibinfo {title} {{What is microbial
  community ecology?}}}\ }\href {https://doi.org/10.1038/ismej.2009.88}
  {\bibfield  {journal} {\bibinfo  {journal} {The ISME Journal}\ }\textbf
  {\bibinfo {volume} {3}},\ \bibinfo {pages} {1223--1230} (\bibinfo {year}
  {2009})}\BibitemShut {NoStop}%
\bibitem [{\citenamefont {Kozubowski}\ and\ \citenamefont
  {Heitman}(2012)}]{Kozubowski2012}%
  \BibitemOpen
  \bibfield  {author} {\bibinfo {author} {\bibnamefont {Kozubowski},
  \bibfnamefont {L.}}and\ \bibinfo {author} {\bibnamefont {Heitman},
  \bibfnamefont {J.}},\ }\bibfield  {title} {\enquote {\bibinfo {title}
  {{Profiling a killer, the development of \textit{Cryptococcus
  neoformans}}},}\ }\href {https://doi.org/10.1111/j.1574-6976.2011.00286.x}
  {\bibfield  {journal} {\bibinfo  {journal} {FEMS Microbiology Reviews}\
  }\textbf {\bibinfo {volume} {36}},\ \bibinfo {pages} {78--94} (\bibinfo
  {year} {2012})},\ \Eprint
  {https://arxiv.org/abs/https://academic.oup.com/femsre/article-pdf/36/1/78/18129553/36-1-78.pdf}
  {https://academic.oup.com/femsre/article-pdf/36/1/78/18129553/36-1-78.pdf}
  \BibitemShut {NoStop}%
\bibitem [{\citenamefont {Lauffenburger}(1991)}]{Lauffenburger1991}%
  \BibitemOpen
  \bibfield  {author} {\bibinfo {author} {\bibnamefont {Lauffenburger},
  \bibfnamefont {D.~A.}},\ }\bibfield  {title} {\enquote {\bibinfo {title}
  {{Quantitative studies of bacterial chemotaxis and microbial population
  dynamics}},}\ }\href {https://doi.org/10.1007/BF02540222} {\bibfield
  {journal} {\bibinfo  {journal} {Microbial Ecology}\ }\textbf {\bibinfo
  {volume} {22}},\ \bibinfo {pages} {175--185} (\bibinfo {year}
  {1991})}\BibitemShut {NoStop}%
\bibitem [{\citenamefont {Liberati}\ \emph {et~al.}(2006)\citenamefont
  {Liberati}, \citenamefont {Urbach}, \citenamefont {Miyata}, \citenamefont
  {Lee}, \citenamefont {Drenkard}, \citenamefont {Wu}, \citenamefont
  {Villanueva}, \citenamefont {Wei},\ and\ \citenamefont
  {Ausubel}}]{Liberati2006}%
  \BibitemOpen
  \bibfield  {author} {\bibinfo {author} {\bibnamefont {Liberati},
  \bibfnamefont {N.~T.}}, \bibinfo {author} {\bibnamefont {Urbach},
  \bibfnamefont {J.~M.}}, \bibinfo {author} {\bibnamefont {Miyata},
  \bibfnamefont {S.}}, \bibinfo {author} {\bibnamefont {Lee}, \bibfnamefont
  {D.~G.}}, \bibinfo {author} {\bibnamefont {Drenkard}, \bibfnamefont {E.}},
  \bibinfo {author} {\bibnamefont {Wu}, \bibfnamefont {G.}}, \bibinfo {author}
  {\bibnamefont {Villanueva}, \bibfnamefont {J.}}, \bibinfo {author}
  {\bibnamefont {Wei}, \bibfnamefont {T.}}, and\ \bibinfo {author}
  {\bibnamefont {Ausubel}, \bibfnamefont {F.~M.}},\ }\bibfield  {title}
  {\enquote {\bibinfo {title} {An ordered, nonredundant library of
  \textit{Pseudomonas aeruginosa} strain pa14 transposon insertion mutants},}\
  }\href {https://doi.org/10.1073/pnas.0511100103} {\bibfield  {journal}
  {\bibinfo  {journal} {Proceedings of the National Academy of Sciences of the
  United States of America}\ }\textbf {\bibinfo {volume} {103}},\ \bibinfo
  {pages} {2833} (\bibinfo {year} {2006})}\BibitemShut {NoStop}%
\bibitem [{\citenamefont {Limoli}\ \emph {et~al.}(2019)\citenamefont {Limoli},
  \citenamefont {Warren}, \citenamefont {Yarrington}, \citenamefont {Donegan},
  \citenamefont {Cheung},\ and\ \citenamefont {O'Toole}}]{Limoli2019}%
  \BibitemOpen
  \bibfield  {author} {\bibinfo {author} {\bibnamefont {Limoli}, \bibfnamefont
  {D.~H.}}, \bibinfo {author} {\bibnamefont {Warren}, \bibfnamefont {E.~A.}},
  \bibinfo {author} {\bibnamefont {Yarrington}, \bibfnamefont {K.~D.}},
  \bibinfo {author} {\bibnamefont {Donegan}, \bibfnamefont {N.~P.}}, \bibinfo
  {author} {\bibnamefont {Cheung}, \bibfnamefont {A.~L.}}, and\ \bibinfo
  {author} {\bibnamefont {O'Toole}, \bibfnamefont {G.~A.}},\ }\bibfield
  {title} {\enquote {\bibinfo {title} {Interspecies interactions induce
  exploratory motility in pseudomonas aeruginosa},}\ }\href
  {https://doi.org/10.7554/eLife.47365} {\bibfield  {journal} {\bibinfo
  {journal} {eLife}\ }\textbf {\bibinfo {volume} {8}},\ \bibinfo {pages}
  {e47365} (\bibinfo {year} {2019})}\BibitemShut {NoStop}%
\bibitem [{\citenamefont {Moran}\ and\ \citenamefont
  {Wernegreen}(2000)}]{Moran2000}%
  \BibitemOpen
  \bibfield  {author} {\bibinfo {author} {\bibnamefont {Moran}, \bibfnamefont
  {N.~A.}}and\ \bibinfo {author} {\bibnamefont {Wernegreen}, \bibfnamefont
  {J.~J.}},\ }\bibfield  {title} {\enquote {\bibinfo {title} {{Lifestyle
  evolution in symbiotic bacteria: Insights from genomics}},}\ }\href
  {https://doi.org/10.1016/S0169-5347(00)01902-9} {\bibfield  {journal}
  {\bibinfo  {journal} {Trends in Ecology \& Evolution}\ }\textbf {\bibinfo
  {volume} {15}},\ \bibinfo {pages} {321--326} (\bibinfo {year}
  {2000})}\BibitemShut {NoStop}%
\bibitem [{\citenamefont {Otterstedt}\ \emph {et~al.}(2004)\citenamefont
  {Otterstedt}, \citenamefont {Larsson}, \citenamefont {Bill}, \citenamefont
  {Ståhlberg}, \citenamefont {Boles}, \citenamefont {Hohmann},\ and\
  \citenamefont {Gustafsson}}]{Otterstedt2004}%
  \BibitemOpen
  \bibfield  {author} {\bibinfo {author} {\bibnamefont {Otterstedt},
  \bibfnamefont {K.}}, \bibinfo {author} {\bibnamefont {Larsson}, \bibfnamefont
  {C.}}, \bibinfo {author} {\bibnamefont {Bill}, \bibfnamefont {R.~M.}},
  \bibinfo {author} {\bibnamefont {Ståhlberg}, \bibfnamefont {A.}}, \bibinfo
  {author} {\bibnamefont {Boles}, \bibfnamefont {E.}}, \bibinfo {author}
  {\bibnamefont {Hohmann}, \bibfnamefont {S.}}, and\ \bibinfo {author}
  {\bibnamefont {Gustafsson}, \bibfnamefont {L.}},\ }\bibfield  {title}
  {\enquote {\bibinfo {title} {Switching the mode of metabolism in the yeast
  \textit{Saccharomyces cerevisiae}},}\ }\href
  {https://doi.org/10.1038/sj.embor.7400132} {\bibfield  {journal} {\bibinfo
  {journal} {EMBO reports}\ }\textbf {\bibinfo {volume} {5}},\ \bibinfo {pages}
  {532--537} (\bibinfo {year} {2004})}\BibitemShut {NoStop}%
\bibitem [{\citenamefont {Pradhan}\ \emph {et~al.}(2022)\citenamefont
  {Pradhan}, \citenamefont {Tanwar}, \citenamefont {Parthasarathy},\ and\
  \citenamefont {Singh}}]{Pradhan2022}%
  \BibitemOpen
  \bibfield  {author} {\bibinfo {author} {\bibnamefont {Pradhan}, \bibfnamefont
  {D.}}, \bibinfo {author} {\bibnamefont {Tanwar}, \bibfnamefont {A.}},
  \bibinfo {author} {\bibnamefont {Parthasarathy}, \bibfnamefont {S.}}, and\
  \bibinfo {author} {\bibnamefont {Singh}, \bibfnamefont {V.}},\ }\bibfield
  {title} {\enquote {\bibinfo {title} {Toroidal displacement of
  \textit{Klebsiella pneumonia} by \textit{Pseudomonas aeruginosa} is a unique
  mechanism to avoid competition for iron},}\ }\href
  {https://doi.org/10.1101/2022.09.21.508880} {\bibfield  {journal} {\bibinfo
  {journal} {bioRxiv}\ ,\ \bibinfo {pages} {2022.09.21.508880}} (\bibinfo
  {year} {2022})}\BibitemShut {NoStop}%
\bibitem [{\citenamefont {Purcell}(1977)}]{Purcell1977}%
  \BibitemOpen
  \bibfield  {author} {\bibinfo {author} {\bibnamefont {Purcell}, \bibfnamefont
  {E.~M.}},\ }\bibfield  {title} {\enquote {\bibinfo {title} {{Life at low
  Reynolds number}},}\ }\href {https://doi.org/10.1119/1.10903} {\bibfield
  {journal} {\bibinfo  {journal} {American Journal of Physics}\ }\textbf
  {\bibinfo {volume} {45}},\ \bibinfo {pages} {3--11} (\bibinfo {year}
  {1977})},\ \Eprint
  {https://arxiv.org/abs/https://pubs.aip.org/aapt/ajp/article-pdf/45/1/3/11809839/3\_1\_online.pdf}
  {https://pubs.aip.org/aapt/ajp/article-pdf/45/1/3/11809839/3\_1\_online.pdf}
  \BibitemShut {NoStop}%
\bibitem [{\citenamefont {Reimer}\ \emph {et~al.}(2021)\citenamefont {Reimer},
  \citenamefont {Sardà Carbasse}, \citenamefont {Koblitz}, \citenamefont
  {Ebeling}, \citenamefont {Podstawka},\ and\ \citenamefont
  {Overmann}}]{Reimer2021}%
  \BibitemOpen
  \bibfield  {author} {\bibinfo {author} {\bibnamefont {Reimer}, \bibfnamefont
  {L.~C.}}, \bibinfo {author} {\bibnamefont {Sardà Carbasse}, \bibfnamefont
  {J.}}, \bibinfo {author} {\bibnamefont {Koblitz}, \bibfnamefont {J.}},
  \bibinfo {author} {\bibnamefont {Ebeling}, \bibfnamefont {C.}}, \bibinfo
  {author} {\bibnamefont {Podstawka}, \bibfnamefont {A.}}, and\ \bibinfo
  {author} {\bibnamefont {Overmann}, \bibfnamefont {J.}},\ }\bibfield  {title}
  {\enquote {\bibinfo {title} {{BacDive in 2022: the knowledge base for
  standardized bacterial and archaeal data}},}\ }\href
  {https://doi.org/10.1093/nar/gkab961} {\bibfield  {journal} {\bibinfo
  {journal} {Nucleic Acids Research}\ }\textbf {\bibinfo {volume} {50}},\
  \bibinfo {pages} {D741--D746} (\bibinfo {year} {2021})},\ \Eprint
  {https://arxiv.org/abs/https://academic.oup.com/nar/article-pdf/50/D1/D741/42058564/gkab961.pdf}
  {https://academic.oup.com/nar/article-pdf/50/D1/D741/42058564/gkab961.pdf}
  \BibitemShut {NoStop}%
\bibitem [{\citenamefont {Rella}\ \emph {et~al.}(2012)\citenamefont {Rella},
  \citenamefont {Yang}, \citenamefont {Gruber}, \citenamefont {Montagna},
  \citenamefont {Luberto}, \citenamefont {Zhang},\ and\ \citenamefont
  {Poeta}}]{Rella2012}%
  \BibitemOpen
  \bibfield  {author} {\bibinfo {author} {\bibnamefont {Rella}, \bibfnamefont
  {A.}}, \bibinfo {author} {\bibnamefont {Yang}, \bibfnamefont {M.~W.}},
  \bibinfo {author} {\bibnamefont {Gruber}, \bibfnamefont {J.}}, \bibinfo
  {author} {\bibnamefont {Montagna}, \bibfnamefont {M.~T.}}, \bibinfo {author}
  {\bibnamefont {Luberto}, \bibfnamefont {C.}}, \bibinfo {author} {\bibnamefont
  {Zhang}, \bibfnamefont {Y.-M.}}, and\ \bibinfo {author} {\bibnamefont
  {Poeta}, \bibfnamefont {M.~D.}},\ }\bibfield  {title} {\enquote {\bibinfo
  {title} {\textit{Pseudomonas aeruginosa} inhibits the growth of
  \textit{Cryptococcus} species},}\ }\href
  {https://doi.org/10.1007/s11046-011-9494-7} {\bibfield  {journal} {\bibinfo
  {journal} {Mycopathologia}\ }\textbf {\bibinfo {volume} {173}},\ \bibinfo
  {pages} {451--461} (\bibinfo {year} {2012})}\BibitemShut {NoStop}%
\bibitem [{\citenamefont {Thormann}, \citenamefont {Beta},\ and\ \citenamefont
  {K\"{u}hn}(2022)}]{Thormann2022}%
  \BibitemOpen
  \bibfield  {author} {\bibinfo {author} {\bibnamefont {Thormann},
  \bibfnamefont {K.~M.}}, \bibinfo {author} {\bibnamefont {Beta}, \bibfnamefont
  {C.}}, and\ \bibinfo {author} {\bibnamefont {K\"{u}hn}, \bibfnamefont
  {M.~J.}},\ }\bibfield  {title} {\enquote {\bibinfo {title} {Wrapped up: The
  motility of polarly flagellated bacteria},}\ }\href
  {https://doi.org/10.1146/annurev-micro-041122-101032} {\bibfield  {journal}
  {\bibinfo  {journal} {Annual Review of Microbiology}\ }\textbf {\bibinfo
  {volume} {76}},\ \bibinfo {pages} {349--367} (\bibinfo {year} {2022})},\
  \bibinfo {note} {pMID: 35650667},\ \Eprint
  {https://arxiv.org/abs/https://doi.org/10.1146/annurev-micro-041122-101032}
  {https://doi.org/10.1146/annurev-micro-041122-101032} \BibitemShut {NoStop}%
\bibitem [{\citenamefont {Trejo-Hern{\'{a}}ndez}\ \emph
  {et~al.}(2014)\citenamefont {Trejo-Hern{\'{a}}ndez}, \citenamefont
  {Andrade-Dom{\'{i}}nguez}, \citenamefont {Hern{\'{a}}ndez},\ and\
  \citenamefont {Encarnaci{\'{o}}n}}]{Trejo-Hernandez2014}%
  \BibitemOpen
  \bibfield  {author} {\bibinfo {author} {\bibnamefont {Trejo-Hern{\'{a}}ndez},
  \bibfnamefont {A.}}, \bibinfo {author} {\bibnamefont
  {Andrade-Dom{\'{i}}nguez}, \bibfnamefont {A.}}, \bibinfo {author}
  {\bibnamefont {Hern{\'{a}}ndez}, \bibfnamefont {M.}}, and\ \bibinfo {author}
  {\bibnamefont {Encarnaci{\'{o}}n}, \bibfnamefont {S.}},\ }\bibfield  {title}
  {\enquote {\bibinfo {title} {{Interspecies competition triggers virulence and
  mutability in \textit{Candida albicans}–\textit{Pseudomonas aeruginosa}
  mixed biofilms}},}\ }\href {https://doi.org/10.1038/ismej.2014.53} {\bibfield
   {journal} {\bibinfo  {journal} {The ISME Journal}\ }\textbf {\bibinfo
  {volume} {8}},\ \bibinfo {pages} {1974--1988} (\bibinfo {year}
  {2014})}\BibitemShut {NoStop}%
\bibitem [{\citenamefont {Turner}, \citenamefont {Souza},\ and\ \citenamefont
  {Lenski}(1996)}]{Turner1996}%
  \BibitemOpen
  \bibfield  {author} {\bibinfo {author} {\bibnamefont {Turner}, \bibfnamefont
  {P.~E.}}, \bibinfo {author} {\bibnamefont {Souza}, \bibfnamefont {V.}}, and\
  \bibinfo {author} {\bibnamefont {Lenski}, \bibfnamefont {R.~E.}},\ }\bibfield
   {title} {\enquote {\bibinfo {title} {Tests of ecological mechanisms
  promoting the stable coexistence of two bacterial genotypes},}\ }\href
  {https://doi.org/https://doi.org/10.2307/2265706} {\bibfield  {journal}
  {\bibinfo  {journal} {Ecology}\ }\textbf {\bibinfo {volume} {77}},\ \bibinfo
  {pages} {2119--2129} (\bibinfo {year} {1996})},\ \Eprint
  {https://arxiv.org/abs/https://esajournals.onlinelibrary.wiley.com/doi/pdf/10.2307/2265706}
  {https://esajournals.onlinelibrary.wiley.com/doi/pdf/10.2307/2265706}
  \BibitemShut {NoStop}%
\bibitem [{\citenamefont {Wadhams}\ and\ \citenamefont
  {Armitage}(2004)}]{Wadhams2004}%
  \BibitemOpen
  \bibfield  {author} {\bibinfo {author} {\bibnamefont {Wadhams}, \bibfnamefont
  {G.~H.}}and\ \bibinfo {author} {\bibnamefont {Armitage}, \bibfnamefont
  {J.~P.}},\ }\bibfield  {title} {\enquote {\bibinfo {title} {{Making sense of
  it all: Bacterial chemotaxis}},}\ }\href {https://doi.org/10.1038/nrm1524}
  {\bibfield  {journal} {\bibinfo  {journal} {Nature Reviews Molecular Cell
  Biology}\ }\textbf {\bibinfo {volume} {5}},\ \bibinfo {pages} {1024--1037}
  (\bibinfo {year} {2004})}\BibitemShut {NoStop}%
\bibitem [{\citenamefont {Wadhwa}\ and\ \citenamefont
  {Berg}(2022)}]{Wadhwa2022}%
  \BibitemOpen
  \bibfield  {author} {\bibinfo {author} {\bibnamefont {Wadhwa}, \bibfnamefont
  {N.}}and\ \bibinfo {author} {\bibnamefont {Berg}, \bibfnamefont {H.~C.}},\
  }\bibfield  {title} {\enquote {\bibinfo {title} {{Bacterial motility:
  Machinery and mechanisms}},}\ }\href
  {https://doi.org/10.1038/s41579-021-00626-4} {\bibfield  {journal} {\bibinfo
  {journal} {Nature Reviews Microbiology}\ }\textbf {\bibinfo {volume} {20}},\
  \bibinfo {pages} {161--173} (\bibinfo {year} {2022})}\BibitemShut {NoStop}%
\bibitem [{\citenamefont {Wu}\ and\ \citenamefont {Berg}(2012)}]{Wu2012}%
  \BibitemOpen
  \bibfield  {author} {\bibinfo {author} {\bibnamefont {Wu}, \bibfnamefont
  {Y.}}and\ \bibinfo {author} {\bibnamefont {Berg}, \bibfnamefont {H.~C.}},\
  }\bibfield  {title} {\enquote {\bibinfo {title} {Water reservoir maintained
  by cell growth fuels the spreading of a bacterial swarm},}\ }\href
  {https://doi.org/10.1073/pnas.1118238109} {\bibfield  {journal} {\bibinfo
  {journal} {Proceedings of the National Academy of Sciences}\ }\textbf
  {\bibinfo {volume} {109}},\ \bibinfo {pages} {4128--4133} (\bibinfo {year}
  {2012})},\ \Eprint
  {https://arxiv.org/abs/https://www.pnas.org/doi/pdf/10.1073/pnas.1118238109}
  {https://www.pnas.org/doi/pdf/10.1073/pnas.1118238109} \BibitemShut {NoStop}%
\bibitem [{\citenamefont {Xiao}\ and\ \citenamefont {Qian}(2000)}]{Xiao2000}%
  \BibitemOpen
  \bibfield  {author} {\bibinfo {author} {\bibnamefont {Xiao}, \bibfnamefont
  {X.}}and\ \bibinfo {author} {\bibnamefont {Qian}, \bibfnamefont {L.}},\
  }\bibfield  {title} {\enquote {\bibinfo {title} {Investigation of
  humidity-dependent capillary force},}\ }\href
  {https://doi.org/10.1021/la000770o} {\bibfield  {journal} {\bibinfo
  {journal} {Langmuir}\ }\textbf {\bibinfo {volume} {16}},\ \bibinfo {pages}
  {8153--8158} (\bibinfo {year} {2000})},\ \Eprint
  {https://arxiv.org/abs/https://doi.org/10.1021/la000770o}
  {https://doi.org/10.1021/la000770o} \BibitemShut {NoStop}%
\bibitem [{\citenamefont {Zengler}\ and\ \citenamefont
  {Zaramela}(2018)}]{Zengler2018}%
  \BibitemOpen
  \bibfield  {author} {\bibinfo {author} {\bibnamefont {Zengler}, \bibfnamefont
  {K.}}and\ \bibinfo {author} {\bibnamefont {Zaramela}, \bibfnamefont
  {L.~S.}},\ }\bibfield  {title} {\enquote {\bibinfo {title} {{The social
  network of microorganisms — how auxotrophies shape complex communities}},}\
  }\href {https://doi.org/10.1038/s41579-018-0004-5} {\bibfield  {journal}
  {\bibinfo  {journal} {Nature Reviews Microbiology}\ }\textbf {\bibinfo
  {volume} {16}},\ \bibinfo {pages} {383--390} (\bibinfo {year}
  {2018})}\BibitemShut {NoStop}%
\bibitem [{\citenamefont {Zhang}, \citenamefont {Turner},\ and\ \citenamefont
  {Berg}(2010)}]{Zhang2010}%
  \BibitemOpen
  \bibfield  {author} {\bibinfo {author} {\bibnamefont {Zhang}, \bibfnamefont
  {R.}}, \bibinfo {author} {\bibnamefont {Turner}, \bibfnamefont {L.}}, and\
  \bibinfo {author} {\bibnamefont {Berg}, \bibfnamefont {H.~C.}},\ }\bibfield
  {title} {\enquote {\bibinfo {title} {The upper surface of an
  \textit{Escherichia coli} swarm is stationary},}\ }\href
  {https://doi.org/10.1073/pnas.0912804107} {\bibfield  {journal} {\bibinfo
  {journal} {Proceedings of the National Academy of Sciences}\ }\textbf
  {\bibinfo {volume} {107}},\ \bibinfo {pages} {288--290} (\bibinfo {year}
  {2010})}\BibitemShut {NoStop}%
\end{thebibliography}%

\newpage
\appendix
\renewcommand{\figurename}{FIG.~S}
\setcounter{figure}{0}    

\begin{figure}
\centering
\includegraphics[width=0.3\textwidth]{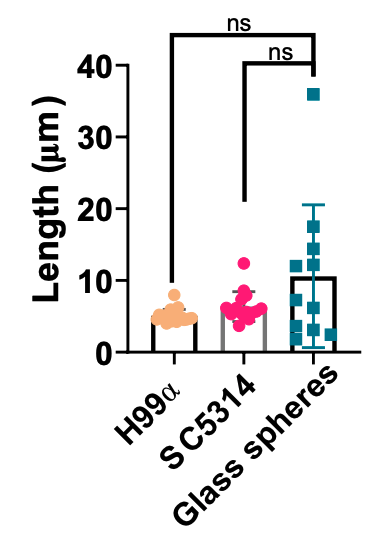}
\caption{Comparative analysis of the length of \textit{C. neoformans} (H99$\alpha$), \textit{C. albicans} (SC5314), and glass spheres, represented as bar plot with mean and standard deviation. Student's t-test with Welch correction was used as a statistical test. The p values are depicted by: ns p>0.05; * p<0.05; ** p<0.01; *** p<0.001; **** p<0.0001.}
\end{figure}


\begin{figure}
\centering
\includegraphics[width=0.6\textwidth]{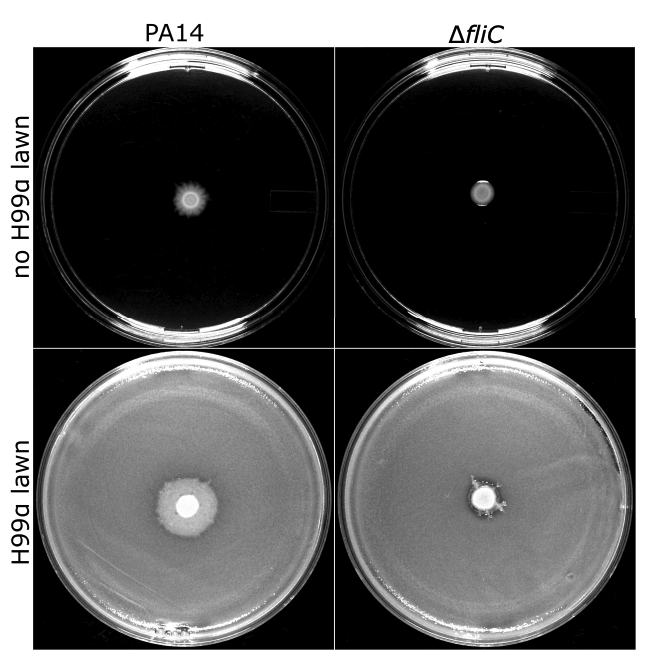}
\caption{Qualitative representation of the colony area occupied by \textit{P. aeruginsoa} (PA14) and flagellum defective \textit{P. aeruginsoa} ($\Delta$\textit{fliC}) in presence and absence of \textit{C. neoformans} (H99$\alpha$) lawn on a 90 mm petri dish. The image was captured following a 32-hour incubation period at a temperature of 25$^{\circ}$C}
\end{figure}

\section*{Effect of nutrient concentration on the spread}
\begin{figure}
\centering
\includegraphics[width=0.5\textwidth]{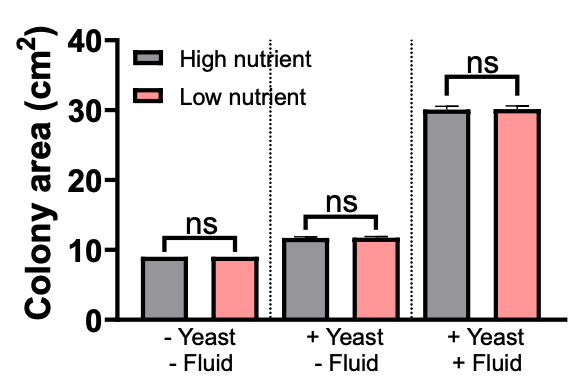}
\caption{The analysis focused on evaluating the model's sensitivity with respect to fluctuations in nutrient content. To assess the potential influence of nutrients on the model, we conducted a comparison of the colony spread area under conditions characterized by high nutrient concentration (10 M) and low nutrient concentration (1 M). A bar plot has been created to visually represent the average and standard error of the mean (SEM) over 100 iterations. The two-way analysis of variance (ANOVA) test is used as a statistical test. The p-values are represented as follows: not significant (ns) when p>0.05, significant (*) when p<0.05, highly significant (**) when p<0.01, very highly significant (***) when p<0.001, and extremely highly significant (****) when p<0.0001.}
\label{fig:nutrientsenstivity}
\end{figure}

To assess the significance of nutrients in the model, we manipulated the starting nutrient concentration and conducted 100 simulations of the model. It was observed that the utilization of two concentrations, specifically 10 M and 1 M, did not result in any significant alteration in the spatial extent of the bacterial colony (see figure~\ref{fig:nutrientsenstivity}). The proposition suggests that the model exhibits independence from perturbations in nutrition availability. 


\begin{figure}
\centering
\includegraphics[width=0.5\textwidth]{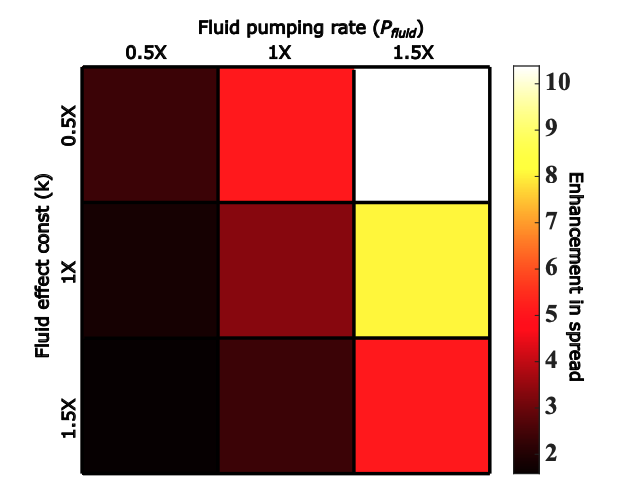}
\caption{The effect of fluid in the spread of a bacteria colony was conducted by simulating the model with variations in the rate of fluid pumping and the constant influence of the fluid (\textit{k}), both in the presence and absence of a yeast agent lawn. The fluid pumping rate and fluid effect constant (\textit{k}) were adjusted by factors of 0.5, 1, and 1.5 relative to their original values. The plot illustrates color tiles, whereby each tile's color corresponds to the average ratio of the area covered by a bacteria colony in the presence and absence of a yeast lawn across 100 instances.}
\label{fig:fluideffect}
\end{figure}

\section*{Effect of fluid pumping by yeast cells}
In order to evaluate the significance of fluid effect and fluid const (\textit{k}), we conducted a simulation where we systematically adjusted the values by a factor of 0.5, 1, and 1.5 in order to quantify the spatial extent of the bacterial colony in the presence and absence of a yeast lawn. It was observed that the spread exhibited a significant rise when the fluid pumping rate rose, while concurrently decreasing the fluid effect constant (\textit{k}) (see figure~\ref{fig:fluideffect}). The observed alteration in the extent of the dispersion area may be attributed to the fact that the presence of fluid is crucial for the spreading of bacteria colony. Thus, the result indicates that the critical factor influencing this process is fluid pumping rate (\textit{P$_{fluid}$}) and fluid effect const (\textit{k}). 

\begin{figure}
\centering
\includegraphics[width=\textwidth]{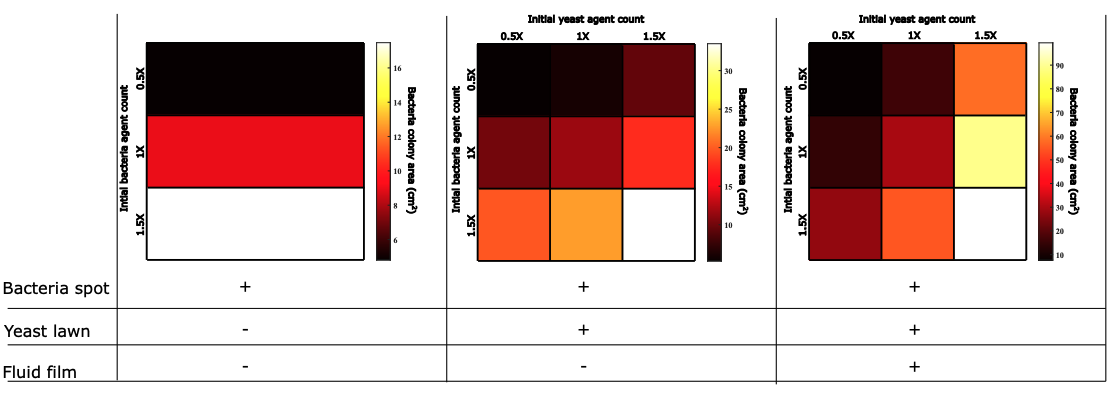}
\caption{The impact of the initial cell count on the propagation of a bacterial colony was estimated. This was achieved by simulating the model with differences in the starting cell count of both bacteria and yeast agents. The initial cell count was varied by the factor of 0.5, 1, and 1.5. These simulations were conducted under three different conditions: the absence of a yeast lawn, the presence of a yeast lawn without fluid pumping, and the presence of a yeast lawn with fluid pumping. The plot shown represents color tiles, where the color of each tile indicates the average area occupied by a colony of bacteria agents throughout a span of 6 hours across 100 different occurrences.}
\label{fig:initialcelldependence}
\end{figure}

\section*{Dependence on the initial cell-loading}
In order to investigate the impact of the starting cell count on the model, we manipulated the initial cell count of both bacteria and yeast agents by applying a scaling factor of 0.5, 1, and 1.5. The model was simulated under three distinct conditions: the absence of yeast, the presence of yeast without fluid pumping, and the presence of yeast with fluid pumping. After conducting a simulation consisting of 100 iterations, it was shown that the spread area of bacteria cells exhibited a positive correlation with the initial cell count when they were placed without a yeast lawn (see figure~\ref{fig:initialcelldependence}). The proliferation of bacterial colonies is enhanced in the presence of a yeast lawn. Moreover, the existence of a fluid coating amplifies the dispersion of bacterial colonies. These results suggest that the initial cell counts play an important role in spreading growing bacteria agents. 

\begin{figure}
\centering
\includegraphics[width=\textwidth]{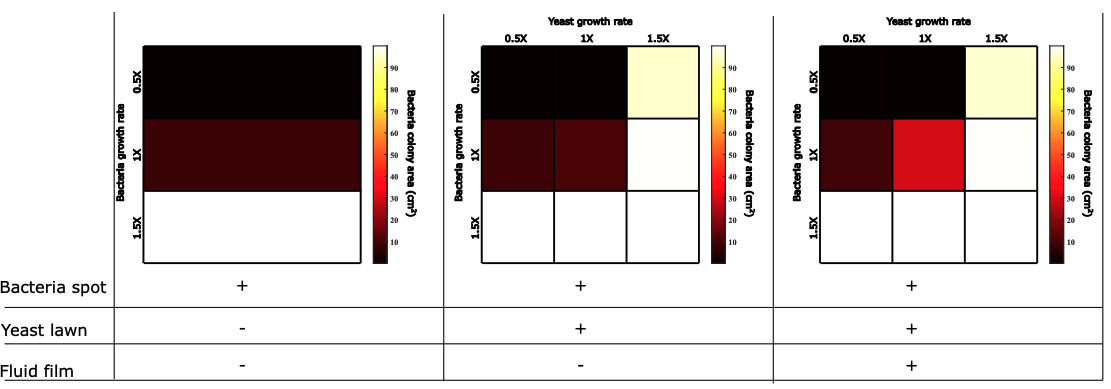}
\caption{The influence of the growth rate on the proliferation of a bacterial colony was estimated. The model incorporated variations in the growth rates of the bacterial and yeast agents. The rate was subject to variations of 0.5, 1, and 1.5. The simulations were performed under three distinct conditions: one without a yeast lawn, another with a yeast lawn but no fluid pumping, and a third with both a yeast lawn and fluid pumping. The presented graphic illustrates color tiles, with each tile's color indicating the average area encompassed by a colony of bacteria agents throughout a duration of 6 hours over a sample of 100 distinct instances.}
\label{fig:growthratedependence}
\end{figure}

\section*{Dependence on the microbe growth rates}
To examine the influence of cell growth rate on the expansion of a bacterial colony, we manipulated the growth rate of both bacterial and yeast agents using scaling factors of 0.5, 1, and 1.5. The simulation was conducted under three unique experimental conditions: the control condition without yeast, the condition with yeast but without fluid pumping, and the condition with yeast and fluid pumping. Based on the results obtained from a simulation comprising 100 iterations, it was shown that the expansion of bacterial cells displayed a positive correlation with their growth rate in the absence of a yeast lawn (see figure~\ref{fig:growthratedependence}). The growth of bacterial colonies is facilitated in the presence of a yeast lawn. Furthermore, the presence of a fluid covering enhances the dispersal of bacterial colonies. The findings of this study indicate that the rate of cellular growth has a significant impact on the dissemination of bacterial agents during their growth process. 



\end{document}